\documentclass[aps,prx,groupedaddress,eqsecnum,preprintnumbers,superscriptaddress,dblfloatfix,nofootinbib,twocolumn]{revtex4-2}
\usepackage[utf8]{inputenc}
\usepackage{hyperref}
\usepackage{graphicx}
\usepackage{amssymb,amsmath}
\usepackage{epstopdf}
\usepackage{aas_macros}
\usepackage{enumitem} 
\usepackage{ulem} 
\usepackage[table,dvipsnames]{xcolor}
\usepackage{bm}
\usepackage{tabularx}
\usepackage{booktabs}
\makeatletter
\newcommand{\customlabel}[2]{%
   \protected@write \@auxout {}{\string \newlabel {#1}{{#2}{\thepage}{#2}{#1}{}} }%
   \hypertarget{#1}{#2}
}
\makeatother

\begin{document}

\title{Halo Formation from Yukawa Forces in the Very Early Universe}

\author{Guillem Dom\`enech} \email{{guillem.domenech}@{itp.uni-hannover.de}}
\affiliation{Institute for Theoretical Physics, Leibniz University Hannover, Appelstraße 2, 30167 Hannover, Germany.}

\author{Derek Inman} \email{derek.inman@ipmu.jp}
\affiliation{Kavli Institute for the Physics and Mathematics of the Universe (WPI), The University of Tokyo Institutes for Advanced Study, The University of Tokyo, Chiba 277-8583, Japan}

\author{Alexander Kusenko} \email{kusenko@ucla.edu}
\affiliation{Kavli Institute for the Physics and Mathematics of the Universe (WPI), The University of Tokyo Institutes for Advanced Study, The University of Tokyo, Chiba 277-8583, Japan}
\affiliation{Department of Physics and Astronomy, University of California, Los Angeles Los Angeles, California, 90095-1547, USA}

\author{Misao Sasaki} \email{{misao.sasaki}@{ipmu.jp}}
\affiliation{Kavli Institute for the Physics and Mathematics of the Universe (WPI), The University of Tokyo Institutes for Advanced Study, The University of Tokyo, Chiba 277-8583, Japan}
\affiliation{Center for Gravitational Physics and Quantum Information, Yukawa Institute for Theoretical Physics, Kyoto University, Kyoto 606-8502, Japan} 
\affiliation{Leung Center for Cosmology and Particle Astrophysics, National Taiwan University, Taipei 10617, Taiwan}

\date{\today}

\begin{abstract}
If long-range attractive forces exist and are stronger than gravity then cosmic halo formation can begin in the radiation-dominated era.  We study a simple realization of this effect in a system where dark matter fermions have Yukawa interactions mediated by scalar particles, analogous to the Higgs boson in the standard model.  We develop a self-consistent description of the system including exact background dynamics of the scalar field, and precise modelling of the fermion density fluctuations.  For the latter, we provide accurate approximations for the linear growth as well as quantitative modelling of the nonlinear evolution using N-body simulations.   We find that halo formation occurs exponentially fast and on scales substantially larger than simple estimates predict.  The final fate of these halos remains uncertain, but could be annihilation, dark stars, primordial black holes, or even the existence of galaxy-sized halos at matter-radiation equality.  More generally, our results demonstrate the importance of mapping scalar-mediated interactions onto structure formation outcomes and constraints for beyond the standard model theories.

\end{abstract}

\preprint{YITP-23-51}
\maketitle

\section{Introduction} 

    The Higgs boson discovery \cite{CMS:2012qbp,ATLAS:2012yve} has confirmed the existence of scalar fields interacting with fermions via Yukawa couplings.  Supersymmetry, axions, string theory and other theories beyond the standard model predict a broad range of new scalar fields, while new fermionic particles could make up the dark matter in the Universe.   We will focus on the long-range effects of the Yukawa interactions, which are important on the length scales shorter than the Compton wavelength of the scalar mediator.  Yukawa forces are attractive, and they can be much stronger than gravity, leading to an instability similar to gravitational collapse in the early universe \cite{Amendola:2017xhl,bib:Flores2021}.

    Obviously, the heavier the scalar field, the shorter is the range of the attractive forces.  However, even for the Higgs boson, there is a window where the Higgs-mediated interactions could be considered long-range on the scale of the horizon in the early universe.   In the case of the Higgs boson during the radiation dominated era, the effective mass at temperature $T\gg 10^2\, {\rm GeV}$ is $m_H(T)\sim m_H(0)+gT\sim gT$. The long-range forces mediated by the Higgs boson are relevant on the distance scales $R=\epsilon H^{-1}$, where $H\sim T^2/M_{\rm Planck}$ and $\epsilon < 1$, if $m_H(T)\sim gT < H/\epsilon $, that is, for temperatures $T> \epsilon g M_{\rm Planck}$.  Any halos that could form under the action of such attractive force would have an upper limit on their masses $M_h< \epsilon M_{\rm Planck}/g^2\sim 10^2 \epsilon M_{\rm Planck}$. In particular, if a black hole were to form from such a halo, it would have the mass smaller than $10^2 M_{\rm Planck}$, and it would quickly evaporate.  However, since a broad variety of scalar fields other than the Higgs are predicted by models of new physics, it is of interest to examine the corresponding instability and growth of perturbations in the early universe.

    Scalar fields are also ubiquitous in cosmology and gravity, from the field responsible for cosmic inflaton to perhaps dark energy and dark matter. The growth of cosmic structures due to additional scalar ``fifth forces'' has been studied in the past, mainly focused on dark matter coupled to a scalar field dark energy \cite{Wetterich:1994bg,Amendola:1999er}, the latter also labelled quintessence \cite{Wetterich:1987fm}. The literature in interacting dark matter and dark energy is vast: for a nice summary of the literature up to 2003, see Section 2 of \cite{Farrar:2003uw}; for a recent review see \cite{Wang:2016lxa}; for a large recollection of references, see \cite{DiValentino:2019jae,Archidiacono:2022iuu}; for early numerical simulations, see \cite{Nusser:2004qu}; and for a coupling to neutrinos instead of dark matter, see \cite{Fardon:2003eh,Amendola:2007yx,Wetterich:2007kr,Ayaita:2011ay,Casas:2016duf,Gogoi:2020qif}.

    In contrast to the late universe, long-range interactions in the early universe could be so large that small primordial density fluctuations collapse to black holes \citep{Amendola:2017xhl}.  However, it is expected that the fermion perturbations will first collapse into virialized objects called halos \citep{bib:Savastano2019} before then forming primordial black holes (PBH) if the halos can efficiently cool \citep{bib:Flores2021}.  A detection of PBHs could therefore be an indication of strong long-range interactions in the early Universe.

    However, in addition to long-range interactions,
    PBHs could come from first order phase transitions \cite{Crawford:1982yz,Kodama:1982sf} and the collapse of supersymmetric  Q-balls \cite{Cotner:2016cvr,Cotner:2019ykd,Flores:2021jas}, but the common assumption is that they form by the collapse of primordial fluctuations, as proposed first by Hawking and Carr \cite{Hawking:1971ei,Carr:1974nx} (see Refs.\cite{Khlopov:2008qy,Sasaki:2018dmp,Carr:2020gox,Green:2020jor,Escriva:2022duf} for recent reviews). PBHs might be a substantial fraction (if not all) of the dark matter \cite{Carr:2020xqk} (and references therein), they could be responsible for some of the LIGO/VIRGO gravitational waves (GWs) events \cite{Bird:2016dcv,Sasaki:2016jop,Wong:2020yig} and might also be the seeds of supermassive black holes \cite{Kawasaki:2012kn,Carr:2018rid}. An appealing aspect of long-range interactions is that, regardless of whether PBHs form, the formation of such early compact structures has a rich phenomenology: they could lead to observable gravitational waves \cite{Flores:2022uzt}, cold electroweak baryogenesis \cite{Flores:2022oef} and magnetogenesis \cite{Durrer:2022cja}. Thus, it is important to explore the formation of such structures in the non-linear regime and clarify their possible collapse to PBHs.

    Most of the literature in long-range interactions in cosmology assume an exponential type coupling to fermionic dark matter, the so-called dilatonic coupling, which is inspired by string theory and scalar-tensor theories of gravity (see, e.g., \cite{Fujii:2003pa,Gubser:2004du}). A standard Yukawa coupling has been considered in, e.g., Refs.~\cite{Farrar:2003uw,Gogoi:2020qif,bib:Flores2021,bib:Domenech2021}. In Ref.~\cite{bib:Domenech2021} new solutions were found in the relativistic fermion regime. In this work, we take a particle physics perspective and consider a renormalizable theory with a scalar field with a quadratic or quartic potential interacting with heavy fermion dark matter via a Yukawa coupling in the very early, radiation dominated, universe.

     We provide below a brief overview of the organization of the paper. In this way, the reader mainly interested in the resulting halo formation might jump directly to the relevant part. We start in \S\ref{sec:motivation} by reviewing the expectation that Yukawa interactions between non-relativistic fermions and a scalar field lead to an exponential growth of fluctuations.  This section also serves as a qualitative orientation for the more detailed calculations in the rest of the paper, which is divided into three main sections:
    \begin{itemize}    
        \item[\S\ref{sec:theory}:] \textbf{Background dynamics.} we present exact solutions for the quadratic (\S\ref{subsec:quadratic}) and quartic (\S\ref{subsec:quartic}) potentials and discuss the implications for general monomial potentials (\S\ref{subsec:powerlaw}). We furthermore describe the parameter space where fermions always remain non-relativistic, which is relevant for the subsequent perturbative and N-body calculations.
        \item[\S\ref{sec:perturbations}:] \textbf{Linear perturbations.}  We show in \S\ref{subsec:instability} that the instability of \S\ref{sec:motivation} also occurs at linear level in the general relativistic setting for a general potential. In \S\ref{subsec:solutionhfl} we specialize to the quartic potential, give the linear solutions and show that Yukawa interactions have a longer range than expected.
        \item[\S\ref{sec:simulation}:] \textbf{N-body simulations.} We first determine the scale free form of the particle equations of motion in \S\ref{subsec:EOM}, and then describe how we evolve them in \S\ref{subsec:nummeth}. The results of our simulations are shown in \S\ref{subsec:haloform}.
    \end{itemize}
    We discuss potential fates of these halos in \S\ref{sec:discussion} and conclude in \S\ref{sec:conclusions}. Details of the calculations and simulations can be found in the appendices. Throughout the paper we assume that the heavy fermions constitute a fraction (or all) of the total dark matter. From now on, we work in natural units where $\hbar=c=1$. A list of the relevant parameters with meaning and definitions used in this paper is provided in appendix \ref{app:elliptic}, Table \ref{tab:1}. 

\section{Motivation: Exponential Growth \label{sec:motivation}}
    To motivate the formation of halos in the radiation era, let us first make a Newtonian fluid analysis of the fermions $\psi$, following the approach of \citep{bib:Wintergerst2010a,bib:Wintergerst2010b,bib:Savastano2019}. This analysis allows us to demonstrate the qualitative features of the model, and especially the exponential growth of structure, before we provide specific details in the following sections. Since there is no creation or annihilation of fermion particles, the mean fermion number density $n_\psi$ in the universe is conserved, that is
    \begin{align}
        \dot{n}_\psi+3Hn_\psi=0\,,
    \end{align}
    where dots are derivatives with respect to cosmic time $t$, i.e.~$\dot{}\equiv d/dt$.  $H=\dot{a}/a$ is the Hubble parameter for a Friedmann–Lemaître–Robertson–Walker (FLRW) metric with scale factor $a(t)$ satisfying
    \begin{align}
        {3}M_{\rm pl}^2H^2=\rho_r+\rho_{m}+\rho_\psi\,,
    \end{align}
    where $M_{\rm pl}^{-2}=8\pi G$ is the reduced Planck mass, $\rho_r\propto a^{-4}$ is the radiation energy density and $\rho_{m}\propto a^{-3}$ is the matter energy density excluding the dark matter fermions. We note that the scalar field energy density may contribute to either radiation or matter, depending on its potential.  In this section we just assume that it does not switch between the two regimes.  We furthermore only consider sufficiently early times such that the dark energy density can be neglected.

    Yukawa interactions between the fermions and the scalar field $\varphi$ lead to a time dependent effective dark matter mass $m_{\rm eff}(t)$ \cite{bib:Flores2021,bib:Domenech2021}, since in general $\varphi=\varphi(t)$. Thus, the fermion energy density, that is $\rho_\psi=m_{\rm eff}n_\psi$ (with $n_\psi\propto a^{-3}$), is not conserved due to exchange with the scalar field, namely
    \begin{align}
        \dot{\rho}_\psi+3H\rho_\psi=B\rho_\psi\,,
    \end{align}
    where $B=\dot{m}_{\rm eff}/m_{\rm eff}$.
    The evolution of the fermion density contrast $\delta_\psi=\delta n_\psi/n_\psi$ is then given by the continuity equation
    \begin{align}
        \dot{\delta}_\psi+\theta_\psi=0 \,, \label{eq:delta1dot}
    \end{align}
    where $\theta_\psi=\vec{\nabla}\cdot\vec{v}_p$ is the divergence of the peculiar velocity $\vec{v}_p$. Due to the time dependent mass, the velocity equation has an extra friction term in addition to Hubble \citep{bib:Wintergerst2010a},
    \begin{align}
        \dot{\theta}_\psi+\left(2H+B\right)\theta_\psi=-\frac{1}{a^2}\nabla^2\phi\,,    
        \label{eq:theta1dot}
    \end{align}
    where $\phi$ represents the potential forces (e.g.~from gravity and the scalar field).  Equations \eqref{eq:delta1dot} and \eqref{eq:theta1dot} can then be combined into a single second order equation, that reads
    \begin{align}
        \ddot{\delta}_\psi+\left(2H+B\right)\dot{\delta}_\psi-\frac{1}{a^2}\nabla^2\phi=0.
        \label{eq:delta2dot}
    \end{align}
    
    Now, we need to specify the potential forces in the system.  The first contribution is from gravity which scales as $\phi_G\propto -1/r$. On subhorizon scales $\phi_G$ satisfies the Poisson equation given by
    \begin{align}
        \frac{1}{a^2}\nabla^2\phi_G = \frac{1}{2}M_{\rm pl}^2 \left[\rho_r\delta_r+\rho_{m}\delta_{m}+\rho_\psi\delta_\psi\right]\,.
        \label{eq:poissonG}
    \end{align}
    Note that, in principle, the gravitational potential includes contributions from both the matter sector and the radiation perturbations in the Universe.  Fortunately, even though it may be the case that $\rho_r\delta_r\gg\rho_{m}\delta_{m}+\rho_\psi\delta_\psi$, cold dark matter effectively only feels self-forces during the radiation era, since the contribution from the radiation (and tightly coupled baryons) averages to zero \citep{bib:Voruz2014}.  Although the dark sector could have many components, we shall assume  for simplicity that only fermions are clustered so that the right-hand side of Eq.~\eqref{eq:poissonG} is dominated by $\rho_\psi\delta_\psi$.  

    The second contribution comes from the scalar field acting as a Yukawa mediator, which typically scales as $\phi_Y \propto \phi_G\exp[-r/\ell]$ \citep{bib:Flores2021}, where $\ell=\ell(t)$ is in general a time dependent length scale of the interaction.  Let us assume the Yukawa force is stronger than gravity by a factor of $2\beta^2\gg1$.  Then, the Yukawa potential is screened relative to the Poisson equation, which leads us to
    \begin{align}
        \frac{1}{a^2}\left[\nabla^2-\ell^{-2}\right]\phi_Y=2\beta^2 \frac{1}{a^2}\nabla^2\phi_G.
    \end{align}
    Setting $\phi=\phi_G+\phi_Y$ and working in Fourier space, we find from \eqref{eq:delta2dot} that 
    \begin{align}
        \ddot{\delta}_\psi+&(2H+B)\dot{\delta}_\psi\nonumber\\=&\frac{3}{2}{H^2}\frac{\rho_\psi}{\rho_r+\rho_m +\rho_\psi}\delta_\psi\left[1+\frac{2\beta^2}{1+(k\ell)^{-2}}\right].
        \label{eq:delta2dot2}
    \end{align}
    We thus see a separation of scales: when $k\ll(\beta\ell)^{-1}$ growth is driven by gravity alone, whereas for $k\gg(\beta\ell)^{-1}$ it is the attractive Yukawa force which dominates.
    
    While the time dependence of $m_{\rm eff}(t)$ and $\ell(t)$ is important, it is instructive to first consider the simple setup where they are constant, e.g., $\dot{m}_{\rm eff}=0$ and $\ell=\bar\ell$, as in \citep{bib:Flores2021}.  Working at early times when dark energy is totally negligible, Eq.~\eqref{eq:delta2dot2} can be converted to a scale-dependent variation of the Meszaros Equation \citep{bib:Meszaros1974}, namely
    \begin{align}
        \delta^{\prime\prime}_\psi+\frac{2+3x}{2x(1+x)}\delta^{\prime}_\psi=\frac{3}{2x(1+x)}\delta_\psi f_\psi\left[1+\frac{2\beta^2}{1+(k\bar\ell)^{-2}}\right]\,, \label{eq:meszaros}
    \end{align}
    where $^\prime=d/dx$ with $x=a/a_{\rm eq}$, $a_{\rm eq}$ being the scale factor when $\rho_r=\rho_m+\rho_\psi$ and $f_\psi=\rho_\psi/(\rho_{m}+\rho_\psi)\le1$.  If we now take the radiation limit $x\ll1$ the general solution to \eqref{eq:meszaros} is given by
    \begin{align}                
        \delta_\psi&=c_1I_0(\sqrt{6\alpha x})+c_2K_0(\sqrt{6\alpha x})\,, \label{eq:deltasoln}\\
        \alpha&=f_\psi\left[1+\frac{2\beta^2}{1+(k\bar\ell)^{-2}}\right]\,,
    \end{align}
    where $I_0$ and $K_0$ are modified Bessel function of order $0$.
    For small arguments, the growing mode is $I_0(\sqrt{6\alpha x})\sim1$ as is traditional in the radiation era, whereas for large arguments it becomes exponential $I_0(\sqrt{6\alpha x})\sim e^{\sqrt{6\alpha x}}/\sqrt{2\pi \sqrt{6\alpha x}}$. 
    We therefore expect that halo formation becomes possible for $6\alpha x\gg1$. This is never possible if gravity is the only force, since $x\ll1$ and $\alpha\sim f_\psi<1$. However, on scales where the Yukawa force is strong, which corresponds to $\alpha\sim2\beta^2{f_\psi}$, an exponential growth can occur for $x\gg1/({12}\beta^2{f_\psi})$.
    We conclude that Yukawa forces could lead to nonlinear evolution and the formation of small halos {\it before} any significant gravitational collapse occurs. 
    
    While this simple picture is appealing, it has neglected the time dependence of $m_{\rm eff}(t)$ and $\ell(t)$ due to the evolution of the scalar field mediator, which can induce significant changes.  For instance, in the cosmologically massless scalar field limit, the system evolves towards minimizing $m_{\rm eff}$ and eventually oscillates around $m_{\rm eff}=0$, which is the exact relativistic fermion regime \citep{bib:Domenech2021}.  While on a time-average fermions remain non-relativistic \citep{bib:Domenech2021}, it is unclear whether the repeated relativistic phases could prevent the Yukawa driven collapse of small scales fluctuations. We are therefore interested in avoiding such regime so that the intuition developed in this section is still valid.  {The next three sections systematically extend the calculation of this section: in \S\ref{sec:theory} we compute exact functional forms of $\ell(t)$ and $m_{\rm eff}(t)$; in \S\ref{sec:perturbations} we solve the linear growth equations including these time dependent parameters; while in \S\ref{sec:simulation} we go beyond this linear theory and compute the fluctuations using N-body simulations.}
    
\section{Background Evolution \label{sec:theory}}

The system under consideration is composed of non-relativistic fermions $\psi$, a scalar field $\varphi$, interacting via a Yukawa coupling with strength $y$, and a radiation fluid with energy density $\rho_r$ and pressure $p_r=\rho_r/3$. We also include a general dark matter sector with energy density $\rho_m$. We take that, for all practical purposes, radiation dominates the energy density of the early universe. Then, assuming that $\psi$ and $\varphi$ are completely decoupled from, or weakly interacting with, the radiation fluid, the Lagrangian of the fermion-scalar system reads
    \begin{align}\label{eq:diraclagrangian}
    {\cal L}({\psi},\varphi)=\bar\psi i\Gamma^\mu D_\mu\psi&-|m_{\psi}+y\varphi|\bar\psi\psi\nonumber\\&-\frac{1}{2}\partial_\mu\varphi\partial^\mu\varphi-V(\varphi)\,,
    \end{align}
    where $V(\varphi)$ is a general potential for the scalar field and we used the chiral symmetry of the fermions to fix the sign of the mass to be positive.
    For later intuition, we note from \eqref{eq:diraclagrangian} that we may interpret the Yukawa interaction as an effective mass for the fermions, namely \begin{equation}\label{eq:meffdefinition}
      m_{\rm eff}=m_\psi+y\varphi\,.
    \end{equation}
    In the action the mass is positive definite and therefore it appears as $|m_{\rm eff}|$. The effective mass \eqref{eq:meffdefinition} is the basic parameter linking the fermions to the scalar field, and one can  express all other quantities, like energy density, pressure, momentum, etc. in terms of it. For more details see \cite{bib:Domenech2021}.

    From now on we approximate the fermions (in thermodynamical equilibrium) as a perfect fluid with energy density $\rho_\psi$ and pressure $p_\psi$. In the non-relativistic limit, the fermion fluid has an energy density given by $\rho_\psi=m_{\rm eff}n_\psi$ and $p_\psi\ll\rho_\psi$ (for more details on the relativistic limit in this set up see \cite{bib:Domenech2021}). In that same limit, it is also possible to identify an effective potential for the scalar field, which is given by
    \begin{equation}\label{eq:Veff}
        V_{\rm eff}=V(\varphi)+\frac{m_{\rm eff}}{|m_{\rm eff}|} \,y\varphi n_\psi\,.
    \end{equation}
   Eq.~\eqref{eq:Veff} leads to the consistent equations of motion in the grand canonical ensemble \cite{bib:Domenech2021}. Note that one may also arrive at \eqref{eq:Veff} from the Lagrangian \eqref{eq:diraclagrangian} by making use of the fermion asymmetry, i.e.~that there are only particles and no antiparticles. In that case, we have that $n_\psi\approx \bar\psi\psi$ (although the correct definition is $n_\psi =\bar \psi \gamma_0 \psi$).

    The background equations of motion are given by the Friedmann equation, energy density conservation and the Klein-Gordon equation. The Friedmann equation reads
    \begin{align}\label{eq:Friedmann}
    3H^2=\rho_r+\rho_{m}+\rho_\psi+\tfrac{1}{2}\dot\varphi^2+V(\varphi)\,.
    \end{align}
    The energy conservation of radiation and dark matter respectively leads to $\rho_r\propto a^{-4}$ and $\rho_m\propto a^{-3}$. The number density conservation implies $n_\psi\propto a^{-3}$. In addition to the Friedmann equation we have the Klein-Gordon equation, namely
    \begin{align}\label{eq:KG}
    \ddot\varphi+3H\dot\varphi+\frac{\partial V_{\rm eff}}{\partial\varphi}=0.
    \end{align}

    From now on, for simplicity, we will refer all quantities with respect to their values at radiation-matter equality, that is when
    \begin{align}\label{eq:rhotoHeq}
    \rho_{ r,\rm eq}=\rho_{m,\rm eq}+\rho_{\psi,\rm eq}=\frac{3}{2}H_{\rm eq}^2M_{\rm pl}^2\,,
    \end{align}
    where $H_{\rm eq}\approx35\,{\rm Mpc}^{-1}\approx 2.2\times 10^{-37}\,{\rm GeV}$. We use the subscript ``eq'' to refer to evaluation at equality. We then rewrite the value of $n_{\psi}$ at equality as
    \begin{align}\label{eq:npsitoHeq}
    n_{\psi, \rm eq}\approx \frac{\rho_{\psi,\rm eq}}{m_\psi}=\frac{f_\psi}{m_\psi}\frac{3}{2}H_{\rm eq}^2M_{\rm pl}^2\,,
    \end{align}
    where we introduced the parameter 
    \begin{align}
    f_\psi\equiv\frac{\rho_{\psi,\rm eq}}{\rho_{m,\rm eq}+\rho_{\psi,\rm eq}}\,,
    \end{align}
    which quantifies the fraction of total dark matter in the form of fermions $\psi$. When radiation dominates the universe, we have from \eqref{eq:Friedmann} that
    \begin{align}
    H(a\ll a_{\rm eq})\approx\frac{H_{\rm eq}}{\sqrt{2}}\left(\frac{a}{a_{\rm eq}}\right)^{-2}\,.
    \end{align}
    The background equations are fully solved once we solve the dynamics of $\varphi$.

    In the cosmologically massless limit of the scalar field, that is when $|\partial V/\partial\varphi|\ll|H\dot\varphi|$, we have that the scalar field grows as \cite{bib:Domenech2021}
    \begin{align}\label{eq:earlysolution}
    \varphi\approx-\frac{2yn_{\psi,\rm eq}}{H_{\rm eq}^2}\frac{a}{a_{\rm eq}}=-\varphi_o\frac{a}{a_{\rm eq}}\,,
    \end{align}
    where we defined for later convenience the parameters
    \begin{align}\label{eq:phio}
    \varphi_o\equiv{3}{\beta f_\psi M_{\rm pl}}\quad{\rm and}\quad\beta\equiv\frac{yM_{\rm pl}}{m_\psi}\,.
    \end{align}
    The parameter $\beta$ is representative of the strength of the Yukawa interaction. The larger the $\beta$, the faster the growth of the scalar field. 
    In \eqref{eq:earlysolution} we take as initial conditions $\varphi(0)=0$. We can do so without loss of generality in the massless regime as a constant value of $\varphi$ can be reabsorbed by a constant shift in $m_{\psi}$. However, when we consider the potential of the scalar field, the value of $\varphi=0$ will in general not coincide with the minimum of $V(\varphi)$. Nevertheless, for analytical simplicity we will later assume that it does.

    Note that by referring all quantities to the time of radiation-matter equality in Eq.~\eqref{eq:earlysolution}, we assumed that the fermions are non-relativistic throughout the entire evolution.\footnote{This would change if the fermions become relativistic. However, the solutions derived are still valid if instead of the time of equality we choose another arbitrary pivot time. The only equations that would not apply are \eqref{eq:rhotoHeq} and \eqref{eq:npsitoHeq}.} This is not necessarily the case as the fermion-scalar system evolves towards the energy minimum which lies at $m_{\rm eff}=m_\psi+y\varphi_c=0$, at which point the fermions become exactly relativistic. In some sense, one may say that the system evolves toward a conformal invariant state where all fields are massless (except for any additional, subdominant, dark matter). In fact, the energy density of fermions and the scalar field also decay as radiation \cite{bib:Domenech2021}. The exact relativistic regime would occur at the critical point given by
    \begin{align}\label{eq:criticalpoint}
    \varphi_c=-\frac{m_\psi}{y}\quad {\rm at}\quad\frac{a_c}{a_{\rm eq}}=\frac{M_{\rm pl}}{\varphi_o\beta}=\frac{1}{3\beta^2f_\psi}\,.
    \end{align}
    After reaching the critical point, the scalar field oscillates around it \cite{bib:Domenech2021}. While it may be possible, it is not clear whether fluctuations can grow during this oscillating regime.  We therefore leave this case for a future work, and instead focus on the non-relativistic fermion regime where exponential growth occurs, as shown in \S~\ref{sec:motivation}.

    We shall be interested in the case when the critical point is never reached and the fermions remain non-relativistic. We then require the condition that:
    \begin{enumerate}[label=(\roman*)]
     \item\label{i} $m_{\rm eff}>0$ at all times, or alternatively $|\varphi|<|\varphi_c|$.
    \end{enumerate}
    However, condition \ref{i} is not sufficient to ensure that fermions remain non-relativistic. To do so, we have two possible conditions:
    \begin{enumerate}[label=(ii\alph*)]
     \item\label{iia}  The fermions are, for some reason, in thermal equilibrium with the radiation bath so we need that $T\ll m_\psi$ in the temperatures of interest. 
     \item \label{iib} The fermions are degenerate and so we need that $m_\psi^3\gg 3\pi^2 n_\psi$.
    \end{enumerate}
    In addition to \ref{iia} or \ref{iib}, we require that:
    \begin{enumerate}[label=(\roman*)]
    \setcounter{enumi}{2}
     \item \label{iii} there are many fermion particles per Hubble volume, that is $n_\psi\gg H^3$. 
    \end{enumerate}
    Condition \ref{iii} is to ensure that later N-body simulations have enough particles inside a Hubble volume. We proceed to derive general bounds on the model parameters.

    First, for condition \ref{i}, a natural way to avoid the $m_{\rm eff}\sim 0$ regime is to consider a sufficiently large mass for the scalar field. In this way, the scalar field's potential might dominate over the Yukawa interaction in \eqref{eq:KG} before reaching $\varphi_c$. A necessary condition is then that the potential satisfies
    \begin{align}\label{eq:conditionpotential}
    |\partial V/\partial\varphi|_{\varphi=\varphi_c}>|yn_\psi|_{\varphi=\varphi_c}\,.
    \end{align}
    There is a large parameter space where this condition is satisfied. For instance, using the early solution \eqref{eq:earlysolution} we find that $|yn_\psi|\propto \varphi^{-3}$. Then, at early times when the scalar field is cosmologically massless we have 
    \begin{align}
    \frac{\partial V_{\rm eff}}{\partial\varphi}\approx\frac{\partial V}{\partial\varphi}-yn_{\psi,\rm eq}\left(\frac{\varphi}{\varphi_o}\right)^{-3}\,.
    \end{align}
    This means that if the potential $V(\varphi)$ increases for increasing $|\varphi|$, there will be a critical value for the parameters of $V(\varphi)$ above which \eqref{eq:conditionpotential} is satisfied. For example, for $V(\varphi)=\tfrac{1}{n}V_o (|\varphi|/\varphi_o)^{n}$ we need $V_o>H_{\rm eq}^2\varphi_o^2$ so that the fermions never reach $m_{\rm eff}=0$.
    
    Second, we may turn conditions \ref{iia},\ref{iib} and \ref{iii} into upper bounds on $\beta$ by requiring that they are satisfied before reaching the critical point $\varphi_c$ \eqref{eq:criticalpoint}. In doing so, we find that, if $f_\psi\sim1$, there is a general upper bound on the value of $\beta$ given by
    \begin{align}\label{eq:upperboundbeta}
     \beta\ll 10^{10}\,.
    \end{align}
    We show the conditions \ref{iia},\ref{iib} and \ref{iii} evaluated at $a=a_c$ in figure \ref{fig:conditiongeneral} and provide the detailed formulas in appendix \ref{app:detailedconditions}.
      
\begin{figure}
\centering
\includegraphics[width=\columnwidth]{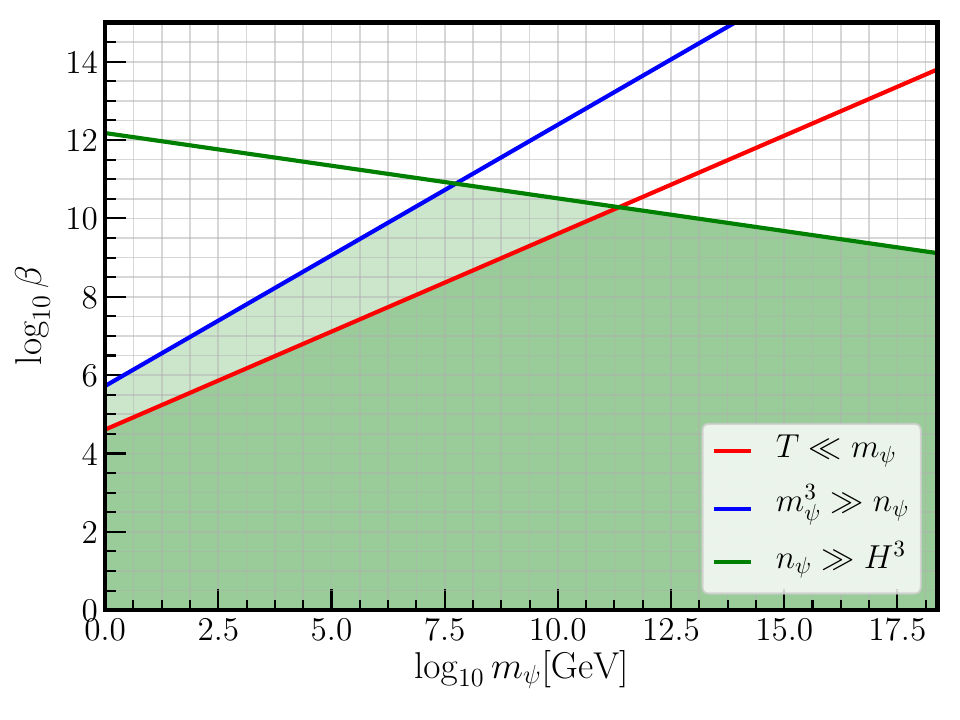}
\caption{Parameter space of $\beta$ in terms of the bare mass of the fermions $m_\psi$. We plot conditions \ref{iia},\ref{iib} and \ref{iii} respectively in red, blue and green. The shaded regions shows the parameter space for $\beta$ such that fermions are non-relativistic, in light and dark green respectively for degenerate and non-degenerate fermions. See how in general $\beta< 10^{10}$.  \label{fig:conditiongeneral}}
\end{figure}

    The evolution of $\varphi$ after the potential dominates depends on the shape of the potential $V(\varphi)$. For this reason, let us consider two typical cases: a quadratic and a quartic potential. As we shall see, while both cases allow for full analytical solutions, the quartic potential turns out to be more suitable for N-body simulations, as the comoving mass of the scalar field, which determines the length scale of the Yukawa force, is independent of the scale factor. We also expect that the quartic term is the dominant contribution to the potential at the high energy scale of the very early universe. We then discuss qualitatively general power-law potentials.

    \subsection{Quadratic potential \label{subsec:quadratic}}

    For a quadratic potential given by $V=\tfrac{1}{2}m_\varphi^2\varphi^2$, the Klein-Gordon equation \eqref{eq:KG} can be simplified to
    \begin{align}\label{eq:KGquadratic}
    \frac{d^2\varphi}{d\xi^2}+\frac{3}{2\xi}\frac{d\varphi}{d\xi}+\varphi+\varphi_*\xi^{-3/2}=0\,.
    \end{align}
    where we defined
    \begin{align}
    \xi \equiv m_\varphi t\quad{\rm and}\quad\varphi_*\equiv\frac{\varphi_o}{2\sqrt{2}}\sqrt{\frac{H_{\rm eq}}{\sqrt{2}m_\varphi}}\,.
    \end{align}
    The particular solution to \eqref{eq:KGquadratic} with initial condition $\varphi(0)=0$ is given by
    \begin{align}\label{eq:solutionquadratic}
    &\varphi=\varphi_*\Big(2^{1/4} \xi^{1/4}
   J_{1/4}(\xi) \Gamma[1/
   4] F\left[\tfrac{1}{4}; \tfrac{3}{4}; \tfrac{5}{4};-\tfrac{\xi^2}{4}\right]\nonumber\\& + 2^{3/4} \xi^{3/4}
   J_{-1/4}(\xi) \Gamma[3/
   4] F\left[\tfrac{1}{2}; \tfrac{5}{4}; \tfrac{3}{2};-\tfrac{\xi^2}{4}\right]\Big)\,,
    \end{align}
    where $J_\nu(x)$ is the Bessel function of order $\nu$, $\Gamma[x]$ is the Gamma function and $F[a;b;c;x]$ is the hypergeometric function. We find that the solution \eqref{eq:solutionquadratic} first grows as \eqref{eq:earlysolution} and then reaches the first maximum at $\xi_1\approx 1.3$ with amplitude $\varphi_1\approx-1.8\varphi_*$. 

    Let us obtain the conditions so that the fermions never reach the relativistic limit in the quadratic case. Here we explicitly write only condition \ref{i}, which yields
    \begin{align}\label{eq:boundquadratic}
    \frac{m_\varphi}{H_{\rm eq}}>2.6f_\psi^2\beta^4\,,
    \end{align}
    and we show all conditions for the quadratic potential in figure \ref{fig:conditionmphi2}. We report the explicit expressions of conditions \ref{iia}, \ref{iib} and \ref{iii} in appendix \ref{app:detailedconditions}.  In figure \ref{fig:conditionmphi2},  we see a vast parameter space where the fermion remains non-relativistic.  

    After reaching the maximum at $\xi_1$, the scalar field decays as $\varphi\propto a^{-3/2}$ and oscillates around $\varphi=0$. Then, the energy density of the scalar field effectively behaves as an additional dark matter component.

\begin{figure}
\centering
\includegraphics[width=0.9\columnwidth]{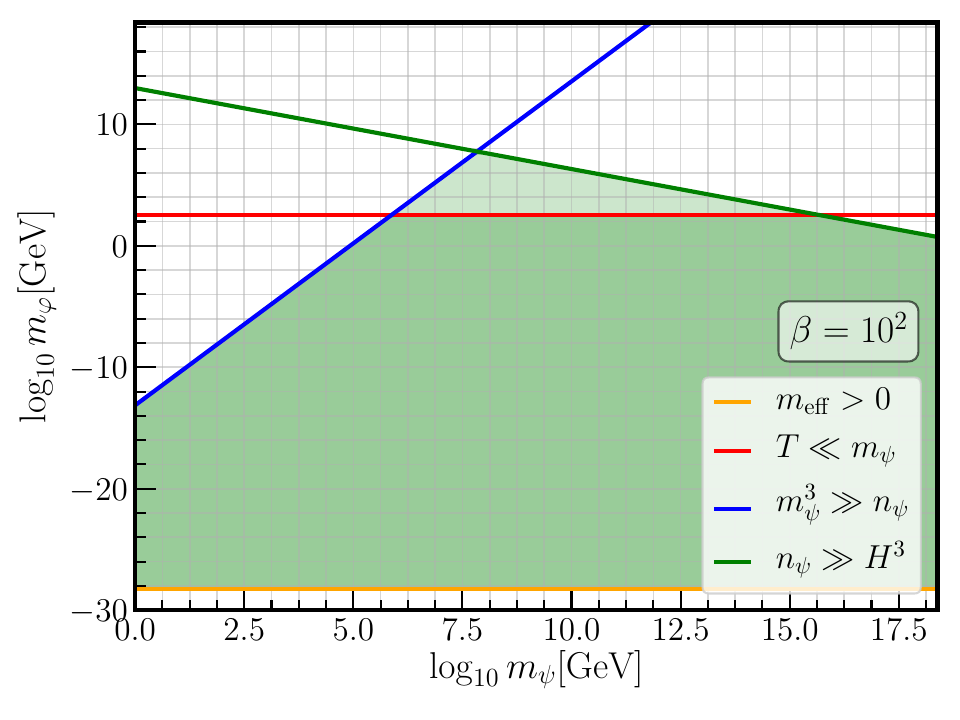}
\caption{Parameter space of $m_\varphi$ in terms of $m_\psi$ for a fixed $\beta=10^2$. We show conditions \ref{i}, \ref{iia}, \ref{iib} and \ref{iii} respectively in orange, red, blue and green. The parameter space where degenerate fermion are non-relativistic is shown with shaded regions, in light and dark green respectively for degenerate and non-degenerate fermions. Note that only the orange line depends on $\beta$ through Eq.~\eqref{eq:boundquadratic}. Thus, increasing the value of $\beta$, rises the orange line and shrinks the parameter space for $m_\varphi$. \label{fig:conditionmphi2}}
\end{figure}

    \subsection{Quartic potential \label{subsec:quartic}}

    In the case of a quartic potential given by $V=\tfrac{1}{4}\lambda\varphi^4$, the Klein-Gordon equation \eqref{eq:KG} can be simplified considerably with the following redefinition:
    \begin{align}\label{eq:ansatzquartic}
    \varphi=\varphi_{\rm eq}\left(\frac{a}{a_{\rm eq}}\right)^{-1}v(a/a_{\rm eq})\,.
    \end{align}
    With the above variable, the Klein-Gordon equation \eqref{eq:KG} reduces to
    \begin{equation}\label{eq:KGquartic}
       \frac{1}{\mu^2}\frac{d^2v}{d(a/a_{\rm eq})^2}+v^3=1\,\quad {\rm where}\quad  \mu^2\equiv\frac{2\lambda\varphi_{\rm eq}^2}{H^2_{\rm eq}}\,,
    \end{equation}
    and we fixed for convenience
    \begin{align}
    \varphi_{\rm eq}\equiv-\left(\frac{y n_{\psi,{\rm eq}}}{\lambda}\right)^{1/3}=-\frac{\varphi_o}{\mu^2}\,.
    \end{align}
    The parameter $\mu^2$ in \eqref{eq:KGquartic} is related to the comoving effective mass of the scalar field by
    \begin{align}
    M_\varphi^2\equiv{a^2V_{\varphi\varphi}}=\frac{3}{2}a_{\rm eq}^2H_{\rm eq}^2\mu^2v^2(a)\,.
    \end{align}
    Also note that by using the value of $\varphi_{\rm eq}$ and $H_{\rm eq}$ we have that \begin{align}\label{eq:mulambda}
    \mu&=\left(\sqrt{2\lambda}\frac{3f_\psi\beta M_{\rm pl}}{H_{\rm eq}}\right)^{1/3}\approx 3.6\times 10^{18}\lambda^{1/6}(f_\psi\beta)^{1/3}\,.
    \end{align}
    From Eq.~\eqref{eq:mulambda} we see that unless $\lambda$ is extremely small, that is $\lambda<5\times 10^{-111}/(f_\psi\beta)^2$, we expect that in general $\mu\gg1$.

    We find that there is an exact solution to \eqref{eq:KGquartic} with the initial condition $\varphi(0)=0$ in terms of elliptic functions.\footnote{Note that, without such initial condition, $v=1$ is also a solution to \eqref{eq:KGquartic}. However, it implies $\varphi=\varphi_c$ far enough in the past.} The solution reads
    \begin{equation}\label{eq:solv}
        v(\zeta)=2^{2/3}\frac{1-{\rm Cn}_\alpha(\zeta)}{\sqrt{3}+1+(\sqrt{3}-1)\,{\rm Cn}_\alpha(\zeta)}\,,
    \end{equation}
    where ${\rm Cn}_\alpha(\zeta)$ is the Jacobian elliptic function
of order $\alpha$ (see appendix \ref{app:elliptic}) and we defined $\zeta\equiv 2^{1/6}\,3^{1/4}\,\mu \,{a}/{a_{\rm eq}}$ and $\alpha\equiv 2^{-3/2}(\sqrt{3}-1)$. The Jacobian elliptic function ${\rm Cn}_\alpha(\zeta)$ is periodic with period equal to $4K_\alpha$ where $K_\alpha$ is the complete elliptic integral of the first kind. In our case we have that $K_\alpha\approx 1.6$.

The solution $\varphi(a)$ \eqref{eq:ansatzquartic} with $v(a)$ given by \eqref{eq:solv}, first grows as \eqref{eq:earlysolution} and reaches a maximum value at $\zeta_{\rm max}\approx2.77$, which corresponds to
\begin{equation}\label{eq:amax}
\varphi_{\rm max}\approx 0.78\,\mu\, \varphi_{\rm eq}\quad{\rm at}\quad a_{\rm max}/a_{\rm eq}\approx1.87\mu^{-1}\,.
\end{equation}
Using \eqref{eq:amax}, we can study the conditions under which the fermions remain non-relativistic. As we did for the quadratic potential, we explicitly write only condition \ref{i}, which yields
\begin{align}\label{eq:boundquartic1}
\lambda> 8\times 10^{-110}f_\psi^4\beta^{10}\,,
\end{align}
and report the details in appendix \ref{app:detailedconditions}. In figure \ref{fig:conditionlambda2} we show all constraints on $\lambda$ and the available parameter space. We see that for $\lambda\sim O(1)$ there is a wide range of parameter space.

\begin{figure}
\centering
\includegraphics[width=\columnwidth]{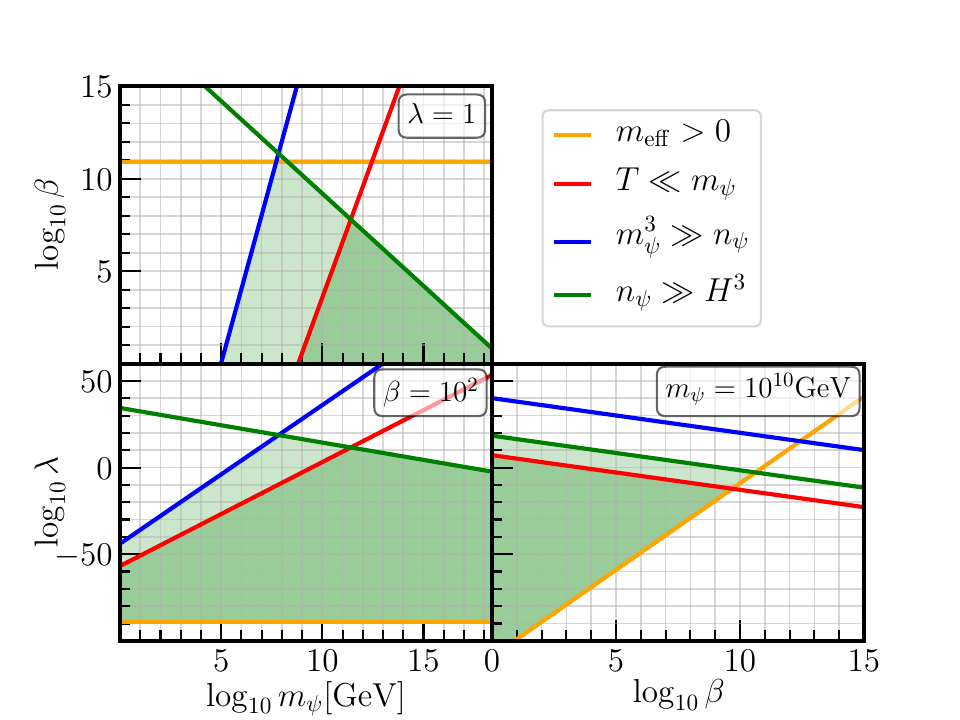}
\caption{Parameter space of $\lambda$, $m_\psi$ and $\beta$ where fermions are non-relativistic. We show in orange, red, blue and green the conditions \ref{i}, \ref{iia}, \ref{iib} and \ref{iii}. The shaded region show the allowed parameter space, in light and dark green respectively for degenerate and non-degenerate fermions. In the top figure we show $\beta$ in terms of $m_\psi$ for $\lambda=1$. Note that, in contrast to the two other panels, the orange line also yields an upper bound for $\beta$. For a fixed value of $m_\psi$, the top figure could also be understood as the parameter space for the Yukawa coupling $y$. In the bottom left panel we have $\lambda$ in terms of $m_\psi$ for $\beta=10^2$. In the bottom right panel, we show $\lambda$ in terms of $\beta$ for a fixed $m_\psi=10^{10}\,{\rm GeV}$.\label{fig:conditionlambda2}}
\end{figure}

After reaching the maximum value $\varphi_{\rm max}$ the scalar field decays as $a^{-1}$ and oscillates. During the oscillations we find that the average value of $v$ and $v^2$ are respectively given by
\begin{align}
{\langle v\rangle}\approx 0.68\quad {\rm and}\quad {\langle v^2\rangle}\approx 0.98\,,
\end{align}
where the brackets refer to oscillation average (defined later in \eqref{eq:averagezeta}).
Note that in contrast to the quadratic potential, the energy density of scalar field now decays like radiation and the oscillations never cross the initial value $\varphi(0)=0$. The latter is due to the fact that, for a quartic potential, the first derivative of $V(\varphi)$ and $yn_\psi$ decay in the same way with the scale factor. Then as $\varphi$ approaches the minimum, the Yukawa interaction starts to dominate again and the system resembles the initial state. Thus, it never crosses the initial condition imposed.

\begin{figure}
\centering
\includegraphics[width=\columnwidth]{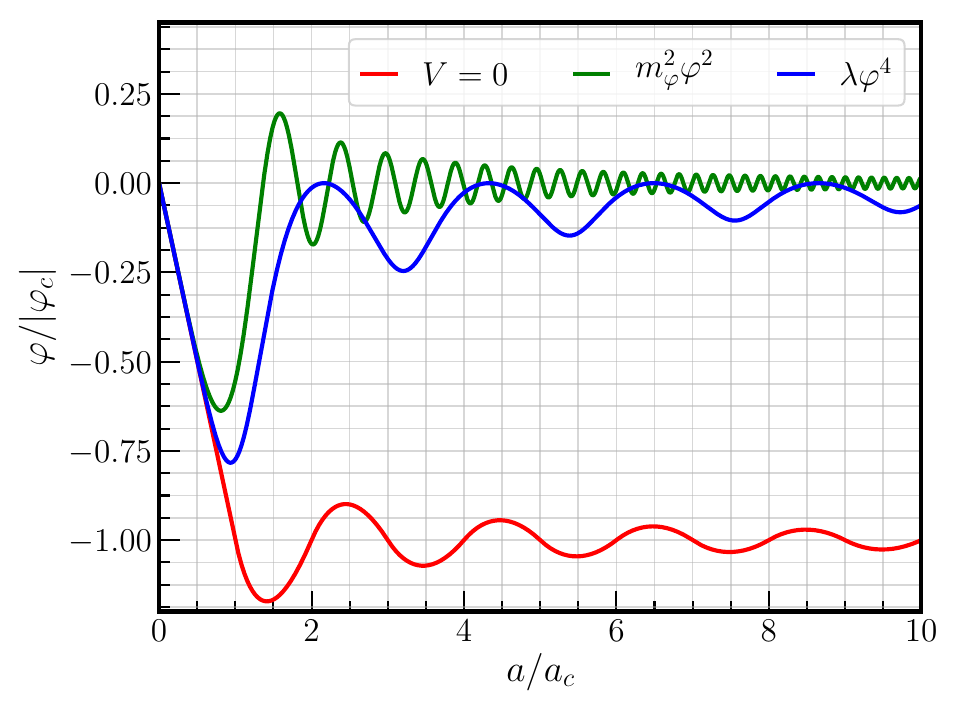}
\caption{Evolution of the scalar field in terms of the scale factor, normalized to the critical value $\varphi_c$ \eqref{eq:criticalpoint}. As an illustrative case we considered $\varphi_o=\beta M_{\rm pl}$, $\varphi_{\rm eq}=-\mu\varphi_o$ and $\varphi_*=2\sqrt{2}\varphi_o$ (for more details see footnote \ref{footnote:generalKG}).  In red, green and blue we respectively show the cases of massless scalar field, quadratic and quartic potentials. All cases share the same early time growth. Then, in the massless case the scalar field oscillates around the critical point. In the other two cases, the potential term dominates before reaching the critical point and the scalar field decays and oscillates.\label{fig:illustrationbackground}}
\end{figure}

In figure \ref{fig:illustrationbackground}, we show an illustration of the behaviour of $\varphi$ in the massless, quadratic and quartic cases. For easier comparison, we chose the values of the parameters such that the coefficients of the Klein-Gordon equation in terms of $v(a)$ are all unity.\footnote{\label{footnote:generalKG}In general, the Klein-Gordon equation including a quadratic and quartic potentials can be rewritten as 
\begin{align}
\frac{d^2v}{da^2}+\frac{\mu^2\varphi_o^2}{\varphi_{\rm eq}^2}v^3+\frac{\varphi_o^4}{64\varphi_*^4}a^2v+{\rm sign}\left[1+\frac{\beta\varphi_o}{M_{\rm pl}}\frac{v}{a}\right]=0\,,
\end{align}
where we now used $\varphi=\varphi_o v(a)/a$. At early enough times with the initial condition $\varphi(0)=0$ we always have that $\varphi\approx -\varphi_o a$\,.
}

    \subsection{Power-law Potential \label{subsec:powerlaw}}

    Let us now discuss the behaviour of an arbitrary monomial following a power-law, namely $V(\varphi)\propto\varphi^{2n}$ with $n\geq 1$. This is possible from an effective field theory point of view. At early times, the system behaves exactly as explained at the beginning of \S\ref{sec:theory}: the scalar field grows due to the Yukawa interaction and decays after the potential dominates. From the effective potential \eqref{eq:Veff}, we expect that the decay of the scalar field either follows the time evolution of the effective minimum of \eqref{eq:Veff} or it oscillates as if no Yukawa interaction is present. In the first case, the scalar field would decay as
    \begin{align}\label{eq:varphimin}
    \varphi_{\rm min}\propto n_\psi^{\frac{1}{2n-1}}\propto a^{-\frac{3}{2n-1}}\,.
    \end{align}
    In the second case, it is known that \cite{Turner:1983he}
    \begin{align}\label{eq:varphinoy}
    \varphi_{\rm no-y}\propto a^{-\frac{3}{n+1}}\,.
    \end{align}
    The subscripts ``min'' and ``no-y'' respectively refer to effective minimum and no Yukawa. The solution which decays slower will dominate the scalar dynamics. After comparing the exponents in \eqref{eq:varphimin} and \eqref{eq:varphinoy}, we see that the oscillations without Yukawa interaction \eqref{eq:varphinoy} dominate for $n<2$, while the field oscillates around the effective minimum \eqref{eq:varphimin} for $n>2$. This also implies that for $n<2$ the oscillations go through $\varphi=0$, while for $n>2$, the field $\varphi$ never returns to the origin (only asymptotically). In the particular case where $n=2$, which corresponds to $\varphi^4$, both exponents are equal and the solution goes back and forth from $\varphi=0$ to $O(\varphi_{\rm min})$.

    From the above background dynamics we can qualitatively understand the growth of perturbations in each case by looking at the comoving effective mass for the scalar field, 
    \begin{align}
    M_\varphi^2=a^2V_{\varphi\varphi}\,.
    \end{align}
    The effective mass determines the Yukawa comoving length scale, namely $\ell=M_{\varphi}^{-1}$, which sets the naive cut-off for the Yukawa force. In other words, not much structure is expected on scales larger than ${\ell}$. Note that we used the word naive as $\ell$ is in general time dependent. Using the solutions \eqref{eq:varphimin} and \eqref{eq:varphinoy}, we see that $\ell$ decreases with $a$ for $n<2$ while it increases for $n>2$. For $n=2$ the amplitude of $\ell$ is independent of $a$. 

    The above discussion implies the following for the growth of perturbations:
    \begin{itemize}
        \item For a quadratic potential ($n=1$), the Yukawa interaction becomes less relevant as the universe expands.  In this case, we expect that halos will have a clear typical mass, or alternatively a peaked mass function as in ref.~\cite{bib:Flores2021} (with some changes in their estimates because the growth stops at some point).
        \item For the quartic potential ($n=2$), one obtains a non-decaying time-oscillating $\ell$. As we shall show later, the time-oscillations of $\ell$ lead to a longer range of the Yukawa interaction (and a larger characteristic mass) than for a constant $\ell$.
        \item For $n>2$, the comoving range of the Yukawa interaction increases with the expansion of the universe. This means that eventually the instability will be of the size of the cosmic horizon. This is also what occurs for dilatonic (exponential) couplings, as in refs.~\cite{Amendola:2017xhl,bib:Savastano2019}, which could be roughly thought as the limit of $n\gg 1$. One must then invoke an additional mechanism to stop the long-range interactions, otherwise fluctuations on all (subhorizon) scales are constantly growing. 
    \end{itemize}

    To anticipate the focus of the simulations, we will eventually specialize on the quartic potential potential with a Yukawa coupling to the fermions because: (1) we expect that the quartic term is the dominant contribution in renormalizable theories in the very early universe and (2) the comoving length scale $\ell$ of the interaction does not redshift with $a$, which allows our simulations to resolve $\ell$ throughout their evolution. 

    In passing, since we are interested in renormalizable theories, we shall find an upper bound on $m_\varphi$ by requiring that the quadratic potential never dominates before $\Delta N$ e-folds after the critical point $a_c$ \eqref{eq:criticalpoint}. Doing this yields 
    \begin{align}
    m_\varphi < 10^{-18} e^{-\Delta N}\,{\rm GeV}\, f_\psi^{4/3}\beta^{7/3}\lambda^{1/6}\,.
    \end{align}
    As we shall later see, $4$ e-folds are sufficient to create non-linear structures. Then, for $\beta\sim 10^{10}$, $\Delta N\sim 4$ and $f_\psi,\lambda\sim O(1)$ we have $m_\varphi<5\,{\rm TeV}$. Smaller $\beta$ or a larger number of e-folds will require a smaller bare mass $m_\varphi$.

\section{Linear Perturbations \label{sec:perturbations}}
 
Having established the background scalar field dynamics, we now investigate the growth of fermion number density fluctuations.  We first show that even though the scalar field exits the cosmologically massless regime at some point, there is still an exponential growth on sufficiently small scales. We then derive and solve the linearized fluid equations governing the fermion perturbations, finding that fluctuations grow on larger than anticipated scales.  We lastly study the limit where the scalar field oscillations rapidly, finding that it converges to a time-independent force law.  We note that while we largely focus on the quartic potential in this section, our justification for the exponential growth is valid for a general potential where the fermions do not enter the relativistic regime.

\subsection{Collapse Instability \label{subsec:instability}}

The Yukawa force is most effective for modes which are subhorizon and below the Yukawa length scale, which is set by the mass of the scalar field. Thus, we expect that a Jeans-like instability appears for wavenumbers such that
\begin{align}
k\gg {\cal H}\quad{\rm and}\quad k^2\gg M_\varphi^2=a^2 V_{\varphi\varphi}\,,
\end{align}
where ${\cal H}=aH$ is the conformal Hubble parameter. Since the exponential growth takes place at small scales and in a short amount of time, compared to the Hubble time, we shall work under the quasi-static approximation, i.e., we neglect time derivatives of the gravitational potential.

To study the perturbations, we work in the Newton gauge in which the metric reads
\begin{align}
ds^2=a^2(-(1+2\Psi)d\eta^2+(1+2\Phi)dx^2)\,,
\end{align}
where $\eta$ is the conformal time, defined by $dt=ad\eta$. We perturb the other variables as $\rho_r\to \rho_r(1+\delta_r)$, $\rho_m\to \rho_m(1+\delta_m)$, $n_\psi\to n_\psi(1+\delta_\psi)$ and $\varphi\to \varphi +\delta\varphi$. In the absence of anisotropic stress, the $i-j$ component of Einstein's equations yields $\Phi+\Psi=0$. See appendix \ref{app:einsteinequations} for more details. For simplicity, we present the equations in terms of e-folds, namely
\begin{align}
dN= {\cal H}d\eta=Hdt=H_{\rm eq}\left(\frac{a}{a_{\rm eq}}\right)^{-2}dt\,.
\end{align}
The number of e-folds $N$ later coincides with what is called the ``superconformal'' time in the N-body simulations.
Then, in the limit where $k\gg {\cal H}$, the gravitational potential $\Phi$ is determined by the Poisson equation, which comes from the $0-0$ component or Hamiltonian constraint, and it is given by
\begin{align}\label{eq:phisubh}
2\frac{ k^2 }{{\cal H}^2}\Phi&=\frac{a^2 }{{\cal H}^2
   M_{\rm pl}^2}(\rho_r\delta_r+\rho_m\delta_m+\rho_\psi\delta_\psi+V_\varphi \delta\varphi) \nonumber\\&+\frac{a^2  \rho_\psi }{{\cal H}^2 M_{\rm pl}^2}\frac{y\delta\varphi}{m_{\rm eff}}+\frac{1}{M_{\rm pl}^2}\frac{d\varphi}{dN}\frac{d\delta \varphi}{dN}\,.
\end{align}
It should be noted that we abused notation with $m_{\rm eff}$ which denotes only the background value of $m_{\rm eff}$.
Also note that in \eqref{eq:phisubh} we have neglected time derivatives of $\Phi$ and we will do so in the subsequent equations. The Klein-Gordon equation for the scalar field fluctuations $\delta\varphi$ reads
\begin{align}\label{eq:dvarphisubh}
\frac{d^2\delta\varphi}{dN^2}&+ \frac{d\delta\varphi}{dN}\nonumber \\&+
   \frac{k^2+a^2V_{\varphi\varphi}}{{\cal H}^2}\delta\varphi+\frac{y }{ m_{\rm eff}}\frac{a^2  \rho_\psi}{{\cal H}^2} \delta_\psi=0\,.
\end{align}
The number density and momentum conservation for the fermion fluids yield
\begin{align}\label{eq:dpsisubh}
\frac{d^2\delta_\psi}{dN^2}+\frac{d\ln |m_{\rm eff}|}{dN}\frac{d\delta_\psi}{dN}-\frac{k^2}{{\cal H}^2}\Phi+\frac{k^2}{{\cal H}^2}\frac{y}{m_{\rm eff}} \delta\varphi=0\,.
\end{align}
The equations of motion for $\delta_r$ and $\delta_m$ are the standard ones (given in appendix \ref{app:einsteinequations}) but, as we shall see, they are not relevant for the following discussion. Also, since we are assuming that $m_{\rm eff}>0$ we drop the absolute value hereon.

We now proceed as follows. With the expectations of an exponential growth of $\delta_\psi$ driven by $\delta\varphi$, we neglect all components in equations \eqref{eq:dvarphisubh} and \eqref{eq:dpsisubh} except for $\delta_\psi$ and $\delta\varphi$. Since the instability leads to an exponential growth for $k^2\gg a^2 V_{\varphi\varphi}$ we also neglect the friction and the potential term in \eqref{eq:dvarphisubh}. Lastly, since we are always in the regime where $|\varphi|<|\varphi_c|$ by condition \ref{i} we take $m_{\rm eff}\approx m_\psi$. We check a posteriori the validity of these assumptions. By doing all the mentioned above, we rewrite \eqref{eq:dvarphisubh} and \eqref{eq:dpsisubh} as
\begin{align}
\frac{d^2}{dN^2}
&
\begin{pmatrix}
\tfrac{\delta\varphi}{M_{\rm pl}}\\
\delta_\psi
\end{pmatrix}
+
\begin{pmatrix}
&\kappa^2&\tfrac{\varphi_o}{M_{\rm pl}}\tfrac{a}{a_{\rm eq}}\\
&\beta\kappa^2&0
\end{pmatrix}
\begin{pmatrix}
\tfrac{\delta\varphi}{M_{\rm pl}}\\
\delta_\psi
\end{pmatrix}=0\,,
\end{align}
where $\varphi_o$ is given by \eqref{eq:phio}, and $\kappa$ is the ratio of the wavenumber to the Hubble parameter, $\kappa\equiv{k}/{\cal H}$. We find that the eigenvalues of the system are given by
\begin{align}
\gamma^2_\pm=\frac{\kappa^2}{2}\left(1\pm\sqrt{1+\frac{4}{\kappa^2}\frac{\varphi_o}{|\varphi_c|}\frac{a}{a_{\rm eq}}}\right)\,,
\end{align}
where we used that $|\varphi_c|=m_\psi/y$ \eqref{eq:criticalpoint}. The eigenvectors are given by
\begin{align}
\vec{e}_\pm=\begin{pmatrix}
\frac{\gamma^2_\pm}{\beta\kappa^2}\\
1
\end{pmatrix}\,.
\end{align}
Note that we always have $\gamma_-^2<0$.  This means that if we neglect the time dependence in $a$ and $\kappa$, there is always an instability and we have that
\begin{align}
\delta_\psi\sim e^{\Gamma N}\,,
\end{align}
where $\Gamma^2\equiv-\gamma_-^2$. Now, let us study two limits. On one hand, for $\kappa^2\gg\tfrac{\varphi_o}{|\varphi_c|}\tfrac{a}{a_{\rm eq}}$ the growth rate is given by
\begin{align}\label{eq:gamma1}
\Gamma^2\approx\frac{\varphi_o}{|\varphi_c|}\frac{a}{a_{\rm eq}}=\frac{a}{a_c}\,.
\end{align}
On the other hand, when $\kappa^2\ll\tfrac{\varphi_o}{|\varphi_c|}\tfrac{a}{a_{\rm eq}}$ the growth rate reads
\begin{align}\label{eq:gamma2}
\Gamma^2\approx \kappa\sqrt{\frac{\varphi_o}{|\varphi_c|}\frac{a}{a_{\rm eq}}}=\kappa\sqrt{\frac{a}{a_c}}\,.
\end{align}

From Eq.~\eqref{eq:gamma1}, we see that $\Gamma\ll1$ for $a<a_c$, namely
during the cosmologically massless regime (see also the discussion around \eqref{eq:criticalpoint}). Eq.~\eqref{eq:gamma2} is not applicable in this regime since it would imply $\kappa\ll1$ but it was only valid on subhorizon scales. 
Thus, although the Yukawa force is effectively a very long-range interaction, it is very weak during the cosmologically massless regime. The physically interesting regime where $\Gamma \gg1$ occurs for $a\gg a_c$ \eqref{eq:criticalpoint} when the scalar field is no longer in the cosmologically massless regime and has non-trivial dynamics. We then conclude that for wavenumbers such that $k\gg{\cal H}$ and $k\gg M_\varphi$ the number density of fermions grow exponentially. 

Before we end this section, let us check that the assumption that $\Phi$, $\delta_r$ and $\delta_m$ are not important for the exponential growth. First, when the exponential growth takes place the scalar field fluctuations are well approximated by
\begin{align}
\frac{\delta\varphi}{M_{\rm pl}}\approx -\frac{\Gamma^2}{\beta\kappa^2}\delta_\psi\,.
\end{align}
This means that for $\kappa\gg\Gamma$ the scalar field fluctuations are subdominant, which is always the case for $a>a_c$ and $k>k_c$, where $k_c={\cal H}_c$. This is an important result for the N-body simulations as it implies that one can treat the Yukawa interaction as an additional force without taking into account the scalar field fluctuations. Second, the requirement that $\Phi$ does not play a role in \eqref{eq:dpsisubh} yields
\begin{align}
\Phi\ll \beta \frac{\delta\varphi}{M_{\rm pl}}=-\frac{\Gamma^2}{\kappa^2}\delta_\psi\,.
\end{align}
Thus, if $\Gamma\ll \kappa$, the amplitude of $\Phi$ is consistently small. This in turn implies from \eqref{eq:phisubh} that $\delta_r, \delta_m\ll \Gamma^2 \delta_\psi$. We see that in the regime where $\kappa\gg1$, $\Gamma\gg1$ and $\Gamma \ll \kappa$ only the fermion fluctuations dominate the cosmological perturbations. In that regime we can thus only consider the fermion fluctuations whose evolution is dictated by an effective potential given by the scalar field.
We can then write
\begin{align}\label{eq:finalfermions}
\frac{d^2\delta_\psi}{dN^2}+\frac{d\ln m_{\rm eff}}{dN}\frac{d\delta_\psi}{dN}=-\frac{k^2}{{\cal H}^2}\phi_Y\,,
\end{align}
and
\begin{align}\label{eq:finalfermions2}
-\frac{k^2+M_\varphi^2}{{\cal H}^2}\phi_Y=\beta^2\frac{a^2 \rho_\psi}{{\cal H}^2M_{\rm pl}^2}\delta_\psi\,,
\end{align}
where we defined
\begin{align}\label{eq:phidefinition}
\phi_Y \equiv \frac{y}{m_{\rm eff}} \delta\varphi\,,
\end{align}
consistent with the notation of \S~\ref{sec:motivation}. Note that because the exponential growth occurs during the oscillations of $\varphi$, the term friction term in \eqref{eq:finalfermions} containing $m_{\rm eff}$ and the effective mass of the scalar field in \eqref{eq:dvarphisubh} could be important.

\subsection{Linear Evolution \label{subsec:solutionhfl}}
From now on, we focus solely on fermion number density fluctuations, since they quickly dominate cosmological perturbations due to their exponential growth on subhorizon scales. We then combine equations \eqref{eq:finalfermions} and \eqref{eq:finalfermions2} into a single equation for $\delta_\psi$, 
    \begin{align} 
        s\frac{d^2\delta_\psi}{ds^2}+\left(1+\frac{d\log m_{\rm eff}}{d\log s} \right)\frac{d\delta_\psi}{ds} = \frac{1}{4}\frac{m_{\rm eff}}{m_\psi}\frac{\delta_\psi}{1+(k\ell)^{-2}}\,, \label{eq:dss}
    \end{align}
where we have introduced a new time variable given by
\begin{align}
s=12\beta^2 f_\psi \frac{a}{a_{\rm eq}} \label{eq:sdef}\,,
\end{align}
so that the system is independent of $\beta$ and $f_\psi$.

Let us now specialise to the quartic potential of \S\ref{subsec:quartic}, where $\phi$ oscillates as a function of $\zeta=\omega s$, as do $m_{\rm eff}$ and $\ell$. 
In terms of the new variable $s$, the frequency of oscillations in $\phi$, given by \eqref{eq:solv}, reads
    \begin{align}\label{eq:omeganumber}
        \omega=\frac{1}{2}\frac{\mu}{2^{5/6} 3^{3/4} f_\psi\beta^2}\approx 4\times 10^{17}\left(\frac{\sqrt{\lambda}}{f^{2}_\psi\beta^{5}}\right)^{1/3}\,,
    \end{align}
which for $f_\psi,\lambda\sim O(1)$ and $\beta\ll 10^{10}$ (see the upper bound \eqref{eq:upperboundbeta}) is in general very large. The frequency $\omega$ enters into the dynamics via the effective mass, that is
    \begin{align}\label{eq:massins}
        \frac{m_{\rm eff}}{m_\psi}=1-\frac{\sqrt{3}}{2s\,\omega^2}\frac{1-{\rm Cn}_\alpha(\omega s)}{\sqrt{3}+1+(\sqrt{3}-1)\,{\rm Cn}_\alpha(\omega s)}\,,
    \end{align}
    and the typical Yukawa length scale, namely
    \begin{align}\label{eq:ellins}
        \ell^{-1}=\bar\ell^{-1}\,\textbf{}3^{1/4}\sqrt{2}\frac{1-{\rm Cn}_\alpha(\omega s)}{\sqrt{3}+1+(\sqrt{3}-1)\,{\rm Cn}_\alpha(\omega s)}\,,
    \end{align}
    where we defined \begin{align}\label{eq:ellbarbar}
    (k_{\rm eq}\bar\ell)^{-1}\equiv2^{-1/3}\,3^{1/4}\mu=6\sqrt{2}f_\psi\beta^2\omega\,,
    \end{align}
    and $k_{\rm eq}={\cal H}_{\rm eq}$.
    With the definition \eqref{eq:ellbarbar} we find that $\langle\ell\rangle\sim 0.8 \,\bar\ell^{-1}$. Thus, we see that while the correction to the mass is suppressed for $\omega\gg1$, yielding ${m_{\rm eff}}\approx{m_\psi}$, the Yukawa length scale $\ell$ varies between $\bar\ell$ and infinity. And, although one might naively conclude that for high frequencies the oscillations average out, the fact that the Yukawa force becomes very long range periodically enhances the power of fluctuations on scales larger than $\bar\ell$.

To obtain initial conditions we note that at early times ($s\ll1$), we have that $m_{\rm eff}/m_\psi\simeq 1-s/8$ and $\bar\ell/\ell\simeq(\omega s)^2/(2\sqrt{2\sqrt{3}})$ and so we can assume that $m_{\rm eff}$ is constant and $\ell\gg k^{-1}$. In this limit we find that
    \begin{align}
        \frac{d\delta_\psi}{ds}\simeq\frac{1}{4}\frac{m_{\rm eff}/m_\psi}{1+d\ln m_{\rm eff}/d\ln s} \frac{\delta_\psi}{1+(k\ell)^{-2}} \simeq \frac{1}{4}\delta_\psi, \label{eq:linearIC}
    \end{align}
    Note that Eq.~\eqref{eq:dss} is precisely the same as Eq.~\eqref{eq:meszaros} and the solution in Eq.~\eqref{eq:linearIC} corresponds to the expansion of the growing mode: $I_0(\sqrt{s\ll1})\simeq 1+s/4$.
    
\begin{figure}
        \centering
        \includegraphics[width=0.9\columnwidth]{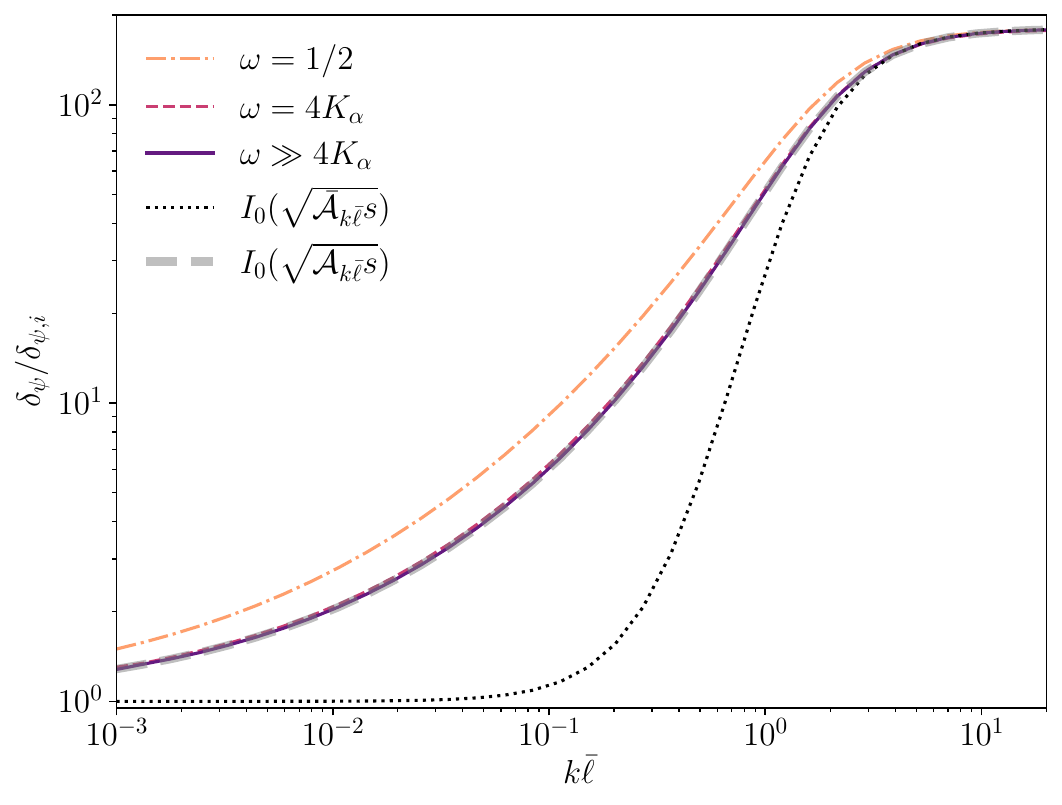}
        \caption{Linear perturbations evaluated at $s=50$ for low ($\omega=1/2$) and high ($\omega=4K_\alpha\sim6.4$ and $\omega=100\gg4K_\alpha$) frequencies.  As the frequency $\omega$ increases the solutions approach the oscillation averaged growing mode solution (dashed grey).  In all cases the oscillations in $\ell$ lead to a substantial increase in large scale perturbations compared to the $\ell=\bar\ell$ case (dotted black).}
        \label{fig:effperturbations}
    \end{figure}

    We numerically solve the linearized fluid equations for the fermion density transfer functions $\delta_\psi/\delta_{\psi,i}$ where we set $\delta_{\psi,i}$ at $s_i=10^{-4}$. We show the resulting transfer functions as a function of $k\bar\ell$ at $s=50$ for $\omega=1/2$, $4K_\alpha(\simeq6.4)$ and $100$ in Fig.~\ref{fig:effperturbations}.  The most dramatic effect is that the force has substantially longer range than in the constant $\ell$ case.  We also find that increasing $\omega$ leads to diminishing changes, and as long as there is roughly a full oscillation ($\omega s \simeq 4 K_\alpha$) before $s=1$ the growth is unchanged. 

\begin{figure}
        \centering
        \includegraphics[width=0.9\columnwidth]{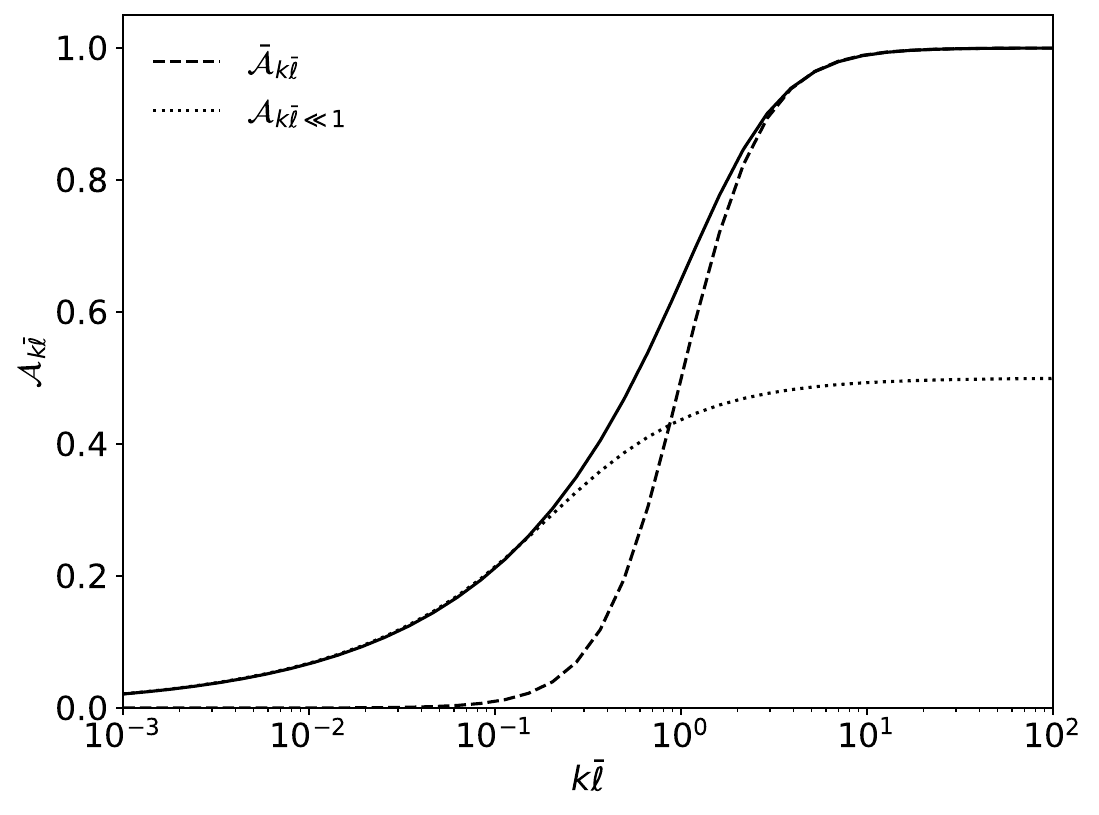}
        \caption{Comparison of the oscillation average  of the Yukawa force law in the high frequency limit (solid), the constant $\ell$ case (dashed) and our approximation for $k\bar\ell\ll1$ (dotted). For comparison, a similar plot for the gravitational force would be a scale independent constant but with an amplitude suppressed by a factor of $2\beta^2\gg1$}
        \label{fig:effpotential}
    \end{figure}

    Now, let us consider the limit when $\omega\gg1$ (or $\zeta\gg1$), which is the general expectation  from \eqref{eq:omeganumber}. In this case, we shall find a good analytical approximation to the solution of \eqref{eq:dss} by replacing highly oscillating background functions by their average over half period, that is
    \begin{align}\label{eq:averagezeta}
        \langle f(\zeta) \rangle &= \frac{1}{2K_\alpha} \int_0^{2K_\alpha} d\zeta_1 f(\zeta_1)\,.
    \end{align}
    We find that $m_{\rm eff}/m_\psi$ quickly approaches unity, and oscillation averages simply lead to $\mathcal{O}(1)$ numbers.
    The more interesting term is the time average of the Yukawa potential that contains $\ell$. As we have that $0<\ell^{-1}\lesssim2\bar\ell^{-1}$, the right hand side in \eqref{eq:dss} oscillates between $1$ and $1/(1+(k\bar\ell)^{-2})$. For later use, let us call this term as
    \begin{align}\label{eq:approx12}
     {\cal A}_{k\bar\ell}\equiv\left\langle\frac{1}{1+(k\ell)^{-2}}\right\rangle\,,
     \end{align}
     while we let $\bar{\cal A}_{k\bar\ell}$ be its evaluation for constant $\ell=\bar\ell$.
     We then note that as we decrease the value of $k\bar\ell$, the term $1/(1+(k\ell)^{-2})$ starts to be mostly negligible except when $\zeta\sim4nK_\alpha$ with $n\in\mathbb{Z}^+$. For $k\bar\ell\ll1$ the term $1/(1+(k\ell)^{-2})$ behaves as a pulse-like function. With this knowledge, we approximate each pulse by a Gaussian, namely we take
     \begin{align}\label{eq:approx1}
     \frac{1}{1+(k\ell)^{-2}}\approx \sum_n \exp\left[-\frac{1}{2\sigma^2}\left(\zeta-4nK_\alpha\right)^2\right]\,.
     \end{align}
     We determine the width of the Gaussian $\sigma$ by finding the time $\zeta$ where ${1}/({1+(k\ell)^{-2}})=e^{-1/2}$. For $k\bar\ell\ll1$, we find that this occurs when $\ell^{-1}\ll1$ and we can use the approximation that $\ell^{-2}\approx \bar\ell^{-2} (\zeta-4nK_\alpha)^4/(8\sqrt{3})$. In this way we find that 
     \begin{align}
     \sigma^2=2^{3/2}3^{1/4}\left(\sqrt{e}-1\right)^{1/2} k\bar\ell \simeq 3 k\bar\ell.
     \end{align}
     We are now ready to compute an analytical approximation for the time average of \eqref{eq:approx1}. After integration we find that
     \begin{align}\label{eq:approx2}
     {\cal A}_{k\bar\ell\ll1}\approx \sqrt{\frac{\pi}{2}}\frac{\sigma}{2K_\alpha}{\rm Erf}\left[\frac{K_\alpha}{\sqrt{2}\sigma}\right]\,.
     \end{align}
     We compare this approximation to the exact oscillation average in Fig.~\ref{fig:effpotential} and find that it is accurate to around $~1.2\%$ for $k\bar\ell\ll1$. Importantly, for $k\bar\ell\ll1$ we find that 
     \begin{align}
     \left\langle\frac{1}{1+(k\ell)^{-2}}\right\rangle_{k\bar\ell\ll1}\propto \sqrt{k\bar\ell}\,,
     \end{align}
     as opposed to the naive $(k\bar\ell)^2$ expectation if $\ell$ were constant. This translates into a longer range interaction and larger power on large scales. We fit the oscillation average with the following asymptotically correct form,
     \begin{align}\label{eq:fit}
        \mathcal{A}_{k\bar\ell} = \left(\frac{1}{1+(k_*/k)^2}\right)^{1/4}
     \end{align}
     with $k_*\bar\ell\simeq2.16$ which is accurate to better than $1.5\%$ regardless of $k\bar\ell$ (simply using $k_*\bar\ell=2$ is accurate to around $5\%$).

    With the time average \eqref{eq:approx12}, we find that Eq.~\eqref{eq:dss} has the same solution as Eq.~\eqref{eq:meszaros}, but with $\mathcal{A}_{k\bar\ell}$ instead of $\bar{\mathcal{A}}_{k\bar\ell}$, namely
     \begin{align}\label{eq:deltapsisol}
     \delta_\psi\approx \delta_{\psi,i}I_0(\sqrt{{\cal A}_{k\bar\ell}s})\,.
     \end{align}
     We show this approximation in Fig.~\ref{fig:effperturbations} as a dashed grey curve, and find that it agrees very well with the computation of the transfer functions at higher frequencies.
     Because ${\cal A}_{k\bar\ell}$ decreases for $k\bar\ell\ll1$ slower than $(k\bar\ell)^{2}$, we find that fluctuations on $k\bar\ell<1$ are significantly enhanced with respect to the case of constant $\ell$. We can also evaluate when a mode $k$ enters the non-linear regime, i.e. when $\delta_\psi(k\bar\ell)\sim1$. We find that this is given by
     \begin{align}\label{eq:snonlinear}
     s_{\rm nl}\approx\frac{1}{4{\cal A}_{k\bar\ell}}W^2\left[-\frac{\delta_{\psi,i}^2}{\pi}\right]\,,
     \end{align}
     where $W(x)$ is the Lambert function of order $-1$.  Let us briefly contextualize how quickly modes $k\gtrsim\bar\ell^{-1}$ become nonlinear.  Based on the analysis of \S\ref{subsec:instability} the fast exponential growth occurs after $a=a_c$ \eqref{eq:criticalpoint} which corresponds to $s(a_c)=4$, and can also be observed in the expansion of Eq.~\eqref{eq:deltapsisol}.  For $\delta_{\psi,i}\sim 10^{-4}$ and $k\bar\ell\gg1$, $s_{\rm nl}\sim 130$ corresponds to just $3.5$ e-folds. Fluctuations become nonlinear very fast. 

     On the other hand, for $k\bar\ell\ll1$ and $\delta_{\psi,i}\ll1$, we have that $|W(x)|$ increases as a combination of $\ln(x)$ and $\ln(\ln(x))$. So a rough order of magnitude estimate would be to replace $W^2(x)\sim \ln^2(x)$, which leads us to
     \begin{align}
     s_{\rm nl}\approx\frac{0.37}{\sqrt{k\bar\ell}}\ln^2\left(\frac{\delta_{\psi,i}^2}{\pi}\right)\,.
     \end{align}
    From this equality, we see that if we require that the mode with $k={\cal H}$ never hits $\delta_\psi=1$ then
     \begin{align}
     s<s_{\rm nl}(k={\cal H})\approx {0.14\,\omega}\ln^4\left(\frac{\delta_{\psi,i}^2}{\pi}\right)\,.
     \end{align} 
    Nevertheless, since $\omega\gg1$ this limit is easily satisfied. We also see that, unless the scalar field decays or fermions collapse to PBHs, the scale $k<k_{\rm eq}$ enters the non-linear regime at $a\sim a_{\rm eq}$ if
    \begin{align}
     \beta\sim2\times 10^{4}\,\left(\frac{\sqrt{\lambda}}{f^5_\psi (k/k_{\rm eq})^3}\right)^{1/11}\ln^{12/11}\left(\frac{\delta_{\psi,i}^2}{\pi}\right)\,.
    \end{align}
    If we take, for example, $\delta_{\psi,i}\sim 10^{-4}$, $k\sim k_{\rm eq}$ and $f_\psi\sim 1$ this implies $\beta\sim6\times 10^5$.

\section{N-body Simulations \label{sec:simulation}}
    Having demonstrated that the fermion density undergoes exponential growth for the quartic potential, our next goal is to study the resulting nonlinear structure that forms from scalar forces, which requires the use of cosmological simulations.  While this scenario is physically quite different from gravitational structure formation, it shares many similarities with gravitational evolution and so can be studied using standard cosmological N-body algorithms, a review of which can be found in \citep{bib:Angulo2022}.  Nonetheless, there are several important differences that need to be taken into account: the simulations are run in the radiation era, rather than matter-dark energy era; the particle mass evolves with time, whereas it is usually constant; and the scalar force has a time varying length scale, rather than the scale-free gravity.  Some aspects of our calculation, such as time stepping criteria and initial conditions, are loosely based on the CUBEP$^3$M code \citep{bib:HarnoisDeraps2013}.  We also use the spherical overdensity halofinder of CUBEP$^3$M, defining halos as having $200\times$ the mean density and at least 100 particles.

    \subsection{Equations of Motion\label{subsec:EOM}}
    Our first goal is to specify the equations of motion that need to be solved in the N-body simulations.  The Hamiltonian for a massive relativistic particle may be written in general as $\textsc{H}=-\frac{1}{2}g^{\mu\nu}p_\mu p_\nu$. However, in the non-relativistic limit where $p_0\gg |\vec{p}|$ we have that the Hamiltonian can also be written, in conformal coordinates, as
    \begin{align}
    {\textsc{H}}\approx \frac{p_0}{a}\approx\frac{\vec{p}^2}{2am_{\rm eff}} + am_{\rm eff}\phi_Y\,,
    \end{align}
    where we used that $-g^{\mu\nu}p_\mu p_\nu=m_{\rm eff}+\delta m_{\rm eff}$ and that $\delta m_{\rm eff}=m_{\rm eff}\phi_Y$. We also dropped the time dependent rest mass $m_{\rm eff}$ and the gravitational potential. In this way we can identify $V_Y\equiv am_{\rm eff}\phi_Y$ as the attractive Yukawa potential. Then Hamilton's equations are given by, 
    \begin{align}
    &\frac{d\vec{x}}{d\eta} \simeq \frac{\vec{p}}{am_{\rm eff}}\,,\\
    &\frac{d\vec{p}}{d\eta} \simeq -am_{\rm eff}\vec{\nabla}\phi_Y \,,
    \end{align}
    where $\eta$ is conformal time, $x$ is a comoving position and $\vec{p}=a m_{\rm eff} \vec{v}_p$ is the conjugate momentum.

    For the simulations, it is more convenient to use equations  that reproduce Newton's laws rather than having friction terms \citep{bib:Pen1998,bib:Martel1998,bib:Angulo2022}.  We therefore utilize the super-conformal time $a dt_N=d\eta$ which in the radiation era ($H=H_r/a^2$, $H_r=a_{\rm eq}^2H_{\rm eq}/\sqrt{2}$) coincides with e-folds $N=H_r t_N$, and define particle velocities as $\vec{v}_N=\vec{p}/m_{\rm eff}$, the potential as $\phi_N=a^2\phi_Y$ and forces via $\vec{f}_N=-m_{\rm eff}\vec{\nabla}\phi_N$.  We then non-dimensionalize the system in the following way:
    \begin{align}
        &t_N=H_r^{-1}t_s\,,\\
        &x=(L/n_c)x_s \leftrightarrow \nabla=(n_c/L)\nabla_s\,,\\
        &v_N=(L H_r/n_c)v_s\,,\\
        &p=(m_\psi L H_r/n_c) p_s \,,\\
        &\phi_N=(L H_r/n_c)^2 \phi_s \,,\\
        &m_{\rm eff}=m_\psi m_s \,,\\
        &\ell=(L/n_c)\ell_s=n_\ell (L/n_c) (\ell/\bar\ell)\,,
    \end{align}
    where $L$ will be the box size of the simulation, $n_c$ is the number of grid cells which will be used in the force calculation, and $n_\ell/n_c=\bar\ell/L$.
    Hamilton's equations are then given by
    \begin{align}
        &\frac{d\vec{x}_s}{dt_s}=\vec{v}_s \label{eq:dxdt_s}\,,\\
        &\frac{d\vec{p}_s}{dt_s}=\vec{f}_s \label{eq:dpdt_s}\,,
    \end{align}
    with $\vec{v}_s=\vec{p}_s/m_s$ and $\vec{f}_s=-m_s\vec{\nabla}_s\phi_s$.  {In these units, Eq.~\eqref{eq:finalfermions2} becomes}
    \begin{align}
        \left(\nabla_s^2-\ell_s^{-2}\right)\phi_s=\frac{s}{4}\delta_\psi \,,\label{eq:poisson_s}
    \end{align}
    {where $s$, defined in Eq.~\eqref{eq:sdef},} can be parameterized as
    \begin{align}
        s=s_ie^{t_s-t_i}.
    \end{align}
    Lastly, {from Eq.~\eqref{eq:massins} we have}
    \begin{align}
        m_s=1-\frac{\sqrt{3}}{2s\,\omega^2}\frac{1-{\rm Cn}_\alpha(\omega s)}{\sqrt{3}+1+(\sqrt{3}-1)\,{\rm Cn}_\alpha(\omega s)} \,,\label{eq:m_s}
    \end{align}
    {while from Eq.~\eqref{eq:ellins}}
    \begin{align}
        \ell_s^{-1}=n_\ell^{-1}\,3^{1/4}\sqrt{2}\frac{1-{\rm Cn}_\alpha(\omega s)}{\sqrt{3}+1+(\sqrt{3}-1)\,{\rm Cn}_\alpha(\omega s)}. \label{eq:ell_s}
    \end{align}
    {Eqs.~\eqref{eq:dxdt_s}-\eqref{eq:ell_s} are the full N-body equations of motion in the Newtonian approximation. It is important to note that the frequency $\omega$ implicitly depends on $L$ since $\omega\propto(k_{\rm eq} \bar\ell)^{-1}$. However, in the high-frequency limit, these equations  no longer depend on the oscillation frequency $\omega$ and they become scale-free (i.e.,~independent of $L$).  The simulations are also invariant to specific values of the fermion mass, $m_\psi$.  The  free parameters that need to be chosen are the initial scale factor $s_i$, the oscillation frequency $\omega$ and the value of the mean Yukawa length $n_\ell$, which in turn set {$\lambda$, $\beta$ and $f_\psi$} in physical parameters.  
    
    \subsection{Numerical Methods \label{subsec:nummeth}}

    We now briefly explain our methodology for the numerical simulations, focusing on differences with respect to standard gravity calculations \citep{bib:Angulo2022}.   Because the particle mass varies with time, we evolve positions and momenta (instead of velocity) according to Eqs.~\eqref{eq:dxdt_s} and \eqref{eq:dpdt_s} using the drift-kick-drift algorithm,
    \begin{align}
        &\vec{x}_s\leftarrow\vec{x}_s+\vec{v}_sdt_s/2 \,,\\
        &\vec{p}_s\leftarrow\vec{p}_s+\vec{f}_sdt_s \,,\\
        &\vec{x}_s\leftarrow\vec{x}_s+\vec{v}_s dt_s/2\,,
    \end{align}
    where we evaluate $\vec{v}_s=\vec{p}_s/m_s(t_s)$ first at $t_s$ and then at $t_s+dt_s$, while $f_s$ is evaluated at $t_s+dt_s/2$.
    Although there are many algorithms to compute $\vec{f}_s$, we have opted to use the simple and robust particle-mesh method \citep{bib:Hockney1988} which involves solving Eq.~\eqref{eq:poisson_s} on a grid.  We first obtain the density contrast $\delta_\psi$ by interpolating particles to a cubic grid of $n_c^3=1024^3$ cells using the Cloud-In-Cell (CIC) method.  We then obtain $\vec{f}_s$ completely in Fourier space via
    \begin{align}
        \vec{f}_s=\frac{i \vec{k}_s}{k_s^2+\ell_s^{-2}}\frac{s}{4}m_s \delta_\psi\,,
        \label{eq:fN}
    \end{align}
    using discrete modes $\vec{k}_s$ chosen to match fourth order accurate finite difference approximations to the gradient and Laplacian.  Having computed the force on the grid, we then interpolate the grid force back to the particles, again using CIC to conserve momentum.  We ensure that the simulation timestep satisfies $dt_s\leq \epsilon_x/{\rm max}[v_s]$ and $dt_s\leq \sqrt{\epsilon_x/{\rm max}[f_s/m_s]}$ where we have set $\epsilon_x$ rather conservatively to be $1/10$ the mean inter-particle distance.  We furthermore also prevent $s$ from changing too rapidly, requiring both $ds$ and $d\ln s$ to be less than $\epsilon_s=0.1$. 

    We next consider how the time varying mass and length scale, Eqs.~\eqref{eq:m_s} and \eqref{eq:ell_s}, affect our calculation.  In this work we consider three different setups: one where we keep $m_s=1$ and $\ell_s=n_\ell$ fixed, one in the low-frequency regime with $\omega=1/2$, and one with $\omega=4K_\alpha$ to realize the high-frequency limit.  When considering the time dependence, $m_s$ is quite straightforward as it only depends on the dimensionless quantities $s$ and $\omega$.  $\ell_s$ on the other hand has a residual dependence on the parameter $n_\ell$.  In other words, even though the system of equations is independent of $L$, the finite resolution of the grid leads to a dependence of our results on the mean length scale.  In Appendix~\ref{app:lbar} we investigate how this potentially affects our results, and have chosen $n_\ell=12$ as a good choice to resolve both wavenumbers larger and smaller than $1/\bar{\ell}$.  In addition to spatial resolution, we also ensure sufficient temporal resolution to track the oscillations in $m_s$ and $\ell_s$ which we do via logarithmic limiters $d\log m_s\le\epsilon_m$ and $d\log\ell_s\le\epsilon_\ell$.  Because $m_s$ simply oscillates near its mean, this condition is straightforward and we set $\epsilon_m=0.1$ throughout, except during the first period where we instead rely on $d\ln s$.  Unfortunately, $\ell_s$ diverges periodically and the logarithmic limiter cannot be maintained.  Instead, we use $\epsilon_\ell=0.1$ while $\ell_s\le n_c$, and simply ensure that $ds$ does not increase until $\ell_s$ returns below $n_c$.

    Lastly, we need to specify the initial conditions of the simulation.  As in Section~\ref{subsec:solutionhfl}, a natural choice of the initial scalefactor, $s_i$, is before Yukawa forces become large, i.e.~when $s_i\ll1$. Because we are considering very small scales that are not at all constrained, we opt for a very simple choice of initial fluctuations: a heuristic scale-invariant spectrum of Gaussian fluctuations with an initial amplitude of $\Delta^2_i=\langle\delta_i^2\rangle=10^{-5}$.  We initially displace $N_p=2\times(n_c/2)^3$ particles from a body-centered-lattice \citep{bib:Joyce2005,bib:Marcos2008} using the Zel'dovich approximation to find the displacement field $-\vec{\nabla}\cdot\vec{\Psi}=\delta_{\psi}(s_i)$ \citep{bib:Zeldovich1970}.  We start the simulation with initial velocities $\vec{V}(s_i)=(1/4)s_i\vec{\Psi}$, which corresponds to the growing mode $\delta_\psi(s)=\delta_\psi(s_i) I_0(\sqrt{s})/I_0(\sqrt{s_i})$.

    As a test of our simulation setup, we compute the dimensionless power spectrum $\Delta^2(k)$ and show it as a function of $k\bar\ell=k_sn_\ell$ for the three oscillation choices in Fig.~\ref{fig:pk_z} at $s=10$ and $s=50$, corresponding to a linear time and just before nonlinearity ensues.  In general we find good agreement on large scales with our linear calculation, while there is some softening on small-scales due to the particle mesh calculation.  We note that at $s=10$ the $\omega=1/2$ case has not completed a full oscillation yet ($s=4K_\alpha/\omega\sim12.8$) and so there is an increase on all scales due to the asymmetric change in $m_s$ and corresponding increase in particle velocity.

    \begin{figure}
        \centering
        \includegraphics[width=0.9\columnwidth]{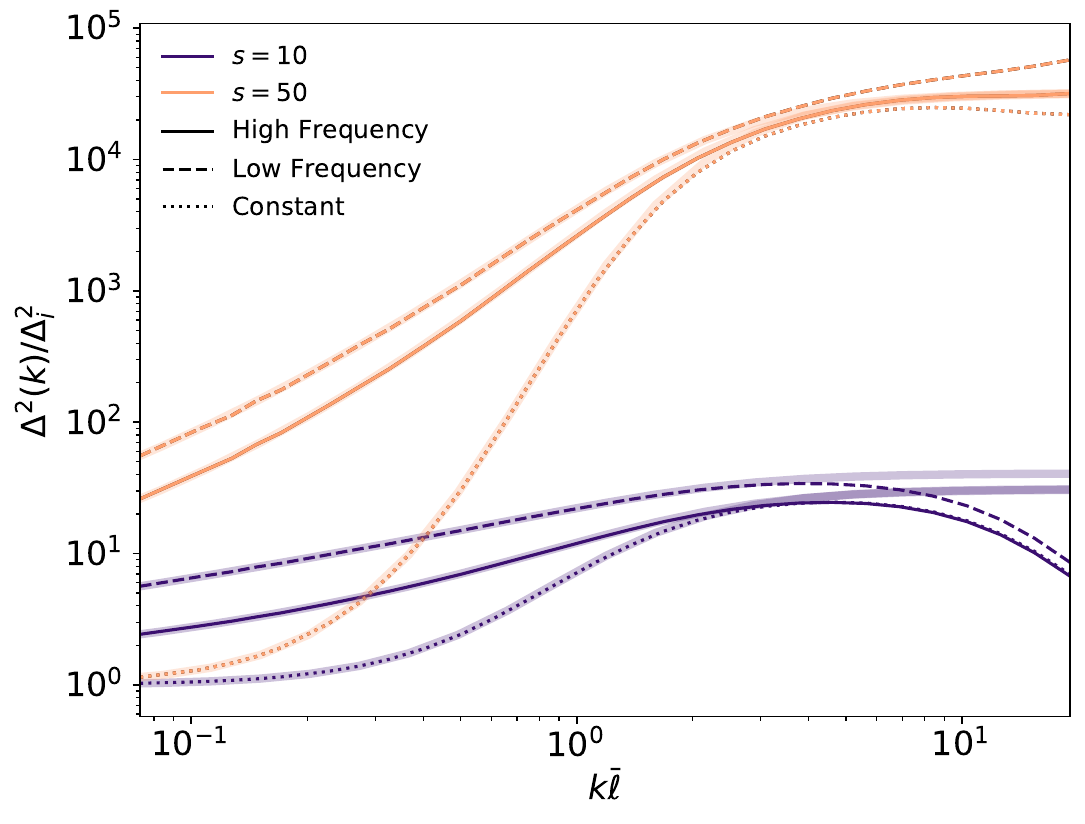}
        \caption{Dimensionless power spectra at $s=10$ and $s=50$ normalized by $\Delta^2_i=10^{-5}$ for the low-frequency, high-frequency and non-oscillating simulations.  The bands around each case are the linear perturbation calculations.}
        \label{fig:pk_z}
    \end{figure}
            
    \subsection{Halo Formation \label{subsec:haloform}}   
    With the cosmological simulations we are now ready to investigate how structure forms under Yukawa forces instead of gravity.  Fig.~\ref{fig:slice_z3} shows a visualization of the fermion density field.  The three rows correspond to the non-oscillating case, the low-frequency ($\omega=1/2$) case and the high-frequency case ($\omega=4K_\alpha\simeq6.4$, which should be representative of the $\omega\rightarrow\infty$ limit).  We see that fragmentation initially resembles that of gravity with distinct web-like features.  However, the finite scalar mass (e.g.~finite $\ell$) leads to a substantially different final state of well-separated halos.  We furthermore see that the final density field, shown at $s=200$ in the right third, is substantially different between the simulations with oscillations and without, with the latter having many small halos throughout, whereas the former has bigger but fewer halos due to the $\propto \sqrt{k\bar\ell}$ scaling.    
    \begin{figure*}
        \centering
        \includegraphics[width=1.0\textwidth]{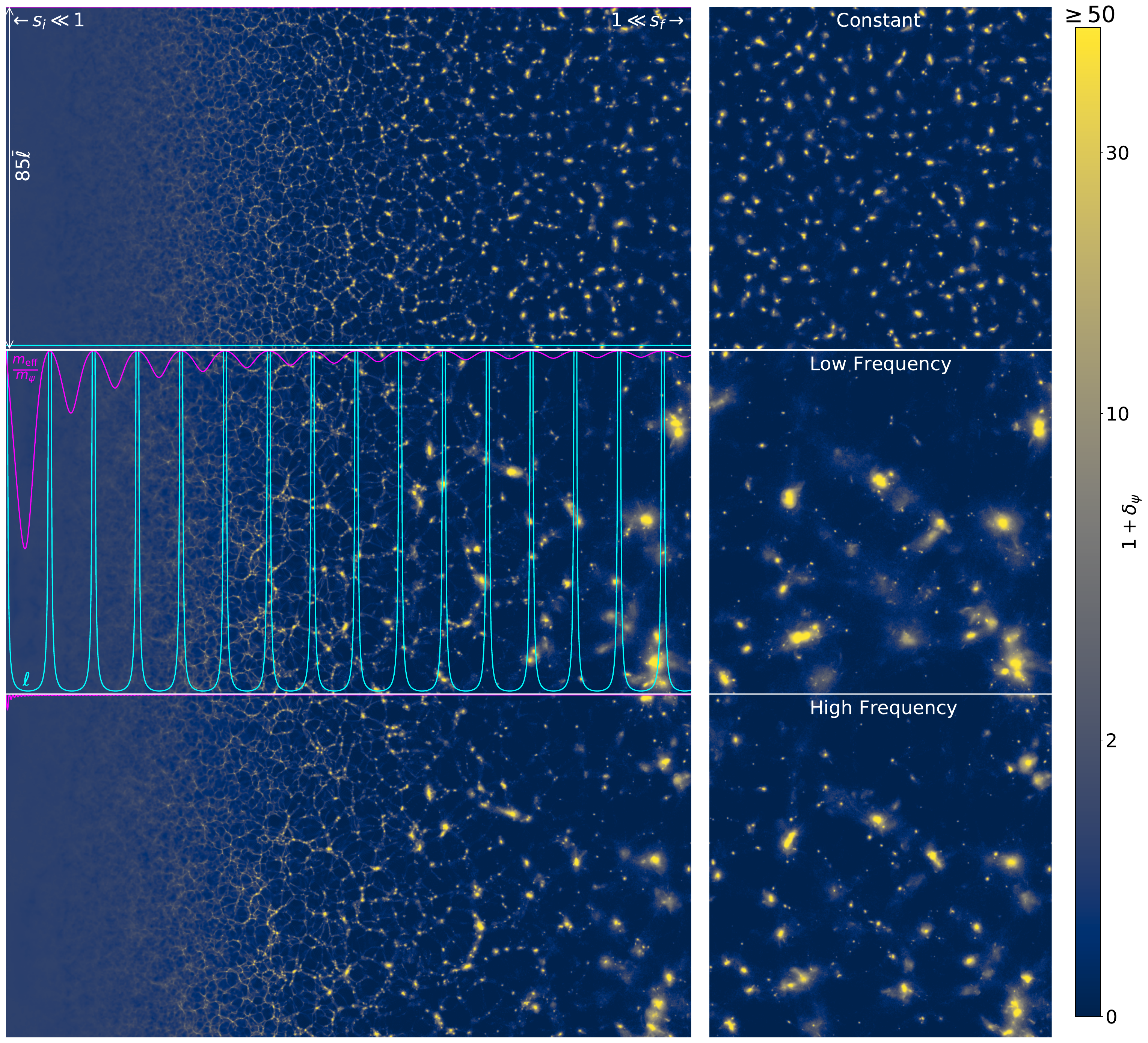}
        \caption{Visualization of fermionic dark matter density field assuming no oscillations (top) and oscillations with $\omega=1/2$ and $4K_\alpha$ (middle and bottom).  The left two-thirds show the time evolution between $s_i=10^{-4}$ and $s_f=200$.  Each pixel column is a density slice equally separated in $s$ and the cubic volume has been periodically wrapped once.  The magenta curve shows the time evolution of the effective mass, $m_{ \rm eff}/m_\psi$, scaled such that 0 is at the bottom and 1 is at the top, while the cyan curve is the time evolution of $\ell$ (omitted in bottom row for clarity).  The right third shows the 2D density field at the final redshift.}
        \label{fig:slice_z3}
    \end{figure*}

    \begin{figure}
        \centering
        \includegraphics[width=0.45\textwidth]{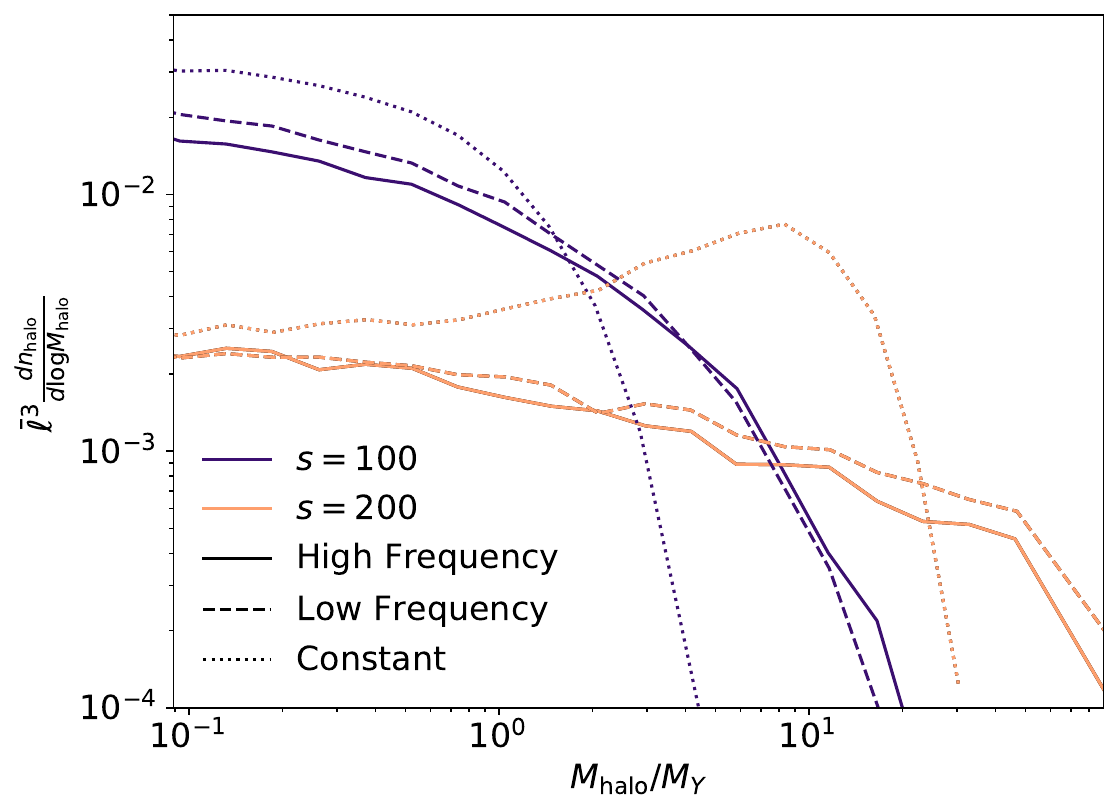}
        \caption{Halo mass function for constant, low frequency and high frequency simulations at $s=100$ and $s=200$. While initially similar, the periodic increases in force range eventually lead to larger halos than the constant case.}
        \label{fig:hmf_z}
    \end{figure}

    To be more quantitative we compute the halo mass function, $dn_{\rm halo}/d\log M_{\rm halo}$ as a function of the mass of the halo, $M_{\rm halo}$.  We normalize masses to
    \begin{align}
    M_{Y}\equiv \frac{4\pi}{3}\rho_{\psi,\rm eq} a_{\rm eq}^3{\bar\ell^3}\,,
    \end{align}
    which is the mass of a halo with an initial radius equal to the Yukawa length scale $a_{\rm eq}\bar\ell$, and corresponds to $(4\pi/3)N_p(n_\ell/n_c)^3\simeq2\times10^3$ N-body particles.
    We show the results in Fig.~\ref{fig:hmf_z} at $s=100$ and $s=200$, which correspond to halfway and completely through the simulation time evolution shown in Fig.~\ref{fig:slice_z3}.  At $s=100$, the halo mass functions are qualitatively similar in all three simulations, with a slight excess/deficit of light/heavy halos in the non-oscillating scenario.  Later however the simulations begin to substantially differ.  The non-oscillating case tends to halos of a similar mass yielding a sharp peak in the mass function.  On the other hand, both oscillating scenarios have a similar halo mass function that extends to much larger scales.

    We now briefly estimate the typical mass of the largest halos as a function of time. We leave a detailed study including radiative cooling as in \cite{bib:Flores2021} for future work. We proceed as follows.  We assume that the largest halos at a given time are those that form when the density fluctuations with wavenumber $k$ enter the non-linear regime, that is when $\delta_{\psi}(k_{\rm nl},s)\sim 1$ \eqref{eq:deltapsisol}. Then, we take that the initial radius of a halo is $O(a/k_{\rm nl})$ and the mass of the halo is given by
    \begin{align}\label{eq:Mk}
    M_{\rm max}(s)&=\frac{4\pi}{3}\rho_{\psi}\left(\frac{a}{k_{\rm nl}(s)}\right)^3= M_{Y} \left({k_{\rm nl}(s)\bar\ell}\right)^{-3}\,,
    \end{align}
    where $k_{\rm nl}(s)$ is the wavenumber that enters the non-linear regime at time $s$ given by inverting Eq.~\eqref{eq:snonlinear}, explicitly
    \begin{align}\label{eq:k(s)}
    k_{\rm nl}(s)\bar\ell=\frac{1}{8 s^2}\frac{W^4\left[-\frac{\delta_{\psi,i}^2}{\pi}\right]}{\sqrt{1-\frac{W^8\left[-\frac{\delta_{\psi,i}^2}{\pi}\right]}{256s^4}}}\,,
    \end{align}
    where we used $k_*\bar\ell\sim 2$ in \eqref{eq:fit}. 
    Using $\bar\ell^{-1}$ from Eq.~\eqref{eq:ellbarbar} we find that
    \begin{align}\label{eq:mellY}
    M_{Y}= \frac{2^{3/2}\pi}{3^{7/4}}\frac{m_\psi}{y\lambda^{1/2}}\approx \frac{6\times 10^{-6}\,{\rm g}}{\beta\sqrt{\lambda}} \,.
    \end{align}
    The physical length scale $\bar\ell_Y$ is given by
    \begin{align}
    \bar\ell_Y=a_{\rm eq}\bar\ell=\frac{2^{-1/6}3^{-1/4}}{(n_{\psi,\rm eq}y\sqrt{\lambda})^{1/3}}\approx \frac{0.23 \,{\rm km}}{(f_\psi\,\beta\,\lambda^{1/2})^{1/3}}.
    \end{align}
Thus, the basic halos are small and light unless $\lambda\ll1$. 

The mass \eqref{eq:mellY} would be the typical mass of the halos if the mass of the scalar field, or alternatively $\ell$, were constant. However, as we have seen in \S\ref{sec:theory}, the actual Yukawa force has a much longer range than $\ell_Y$ due to the time-oscillations of the mass. This causes the halos to keep merging constantly over time until the long-range interaction stops. For example, let us consider that $\varphi$ has a tiny mass $m_\varphi$ so that the quadratic term does not play a role in the early evolution of the system but eventually dominates the potential. Once this occurs, the comoving Yukawa length scale decreases with the scale factor, that is $\ell= (a m_\varphi)^{-1}$ without oscillations, quickly shrinking the range of the interaction. Then as an extreme case we take that $m_\varphi\sim H_{\rm eq}$, like ultra-light dark matter, and we can roughly evaluate the non-linear scale \eqref{eq:k(s)} at radiation-matter equality, where $s=12f_\psi\beta^2$, with $\delta_{\psi,i}\sim 10^{-4}$, which gives
\begin{align}
k_{\rm nl}(s=12f_\psi\beta^2)\bar\ell\sim \frac{230}{f_\psi^2\beta^4}\,.
\end{align}
Then, the maximum halo mass \eqref{eq:Mk} is roughly
 \begin{align}\label{eq:Mk2}
    M_{\rm max}(s=12f_\psi\beta^2)&\approx 5\times 10^{42}\,{\rm g}\,\frac{f_\psi^6}{\sqrt{\lambda}}\left(\frac{\beta}{10^5}\right)^{11}\,.
\end{align}
Thus, as time goes on, tiny halos merge several times and the resulting halos are huge, even by cosmological standards as $5\times10^{42}{\rm\ g}\sim2.5\times10^9 M_\odot$.  The maximum halo mass only depends on the time where the long-range forces disappears by, e.g., the decay of the scalar field or the formation of PBHs. The physical size of the halo can be estimated to be
\begin{align}\label{eq:Rhalo}
    R_{\rm max}(a_0)&\sim\left(\frac{M_{\rm max}}{4\pi\Delta\rho_\psi/3}\right)^{1/3}=\frac{1+z_{\rm eq}}{\Delta ^{1/3}k_{\rm nl}\bar\ell}\bar\ell_Y\nonumber\\&\approx 40\,{\rm kpc}\frac{f_\psi^{5/3}}{\lambda^{1/6}}\left(\frac{\beta}{10^5}\right)^{11/3}\,.
\end{align}
where we used $\Delta\sim 200$ in the last step.

Below we list three examples in which the parameters in the model enter the allowed range for the validity of N-body simulations discussed in \S\ref{sec:theory}:
\begin{itemize}
\item \textit{Example 1}: we require $O(0.1)$ values for the Yukawa and quartic coupling; $\beta=10^7$, $m_\psi= 10^{11}{\rm GeV}$, $y=0.4$ and $\lambda=0.3$. This yields a small Yukawa scale and basic halo mass with  $\bar\ell_Y=1\,\rm m$, $M_{Y}=10^{-10}\,{\rm g}$.
\item \textit{Example 2}: we allow only for small values of the quartic coupling and aim for large $\beta$;
$\beta=10^9$, $m_\psi= 5\times 10^{8}{\rm GeV}$, $y=0.2$ and $\lambda=10^{-19}$. This time the scale and mass is slightly larger than in example 1 with $\bar\ell_Y=0.4\,\rm km$, $M_{Y}=4\,{\mu\rm g}$.
\item \textit{Example 3}: we aim for a large halo mass allowing for tiny couplings; $\beta=10^5$, $m_\psi=10^{6}{\rm GeV}$, $y=4\times 10^{-8}$ and $\lambda=10^{-50}$. Now we have a large and massive basic halo with $\bar\ell_Y=10^6\,\rm km$, $M_{Y}=10^{17}\,{\rm g}$.
\end{itemize}
The maximum halo mass depends on the concrete evolution of the scalar field until equality, which is beyond the validity of our approximations. Let us emphasize that the constraints on the parameter space, presented in \S\ref{sec:theory} and in Appendix \ref{app:detailedconditions}, used in the above examples only concern the validity of the N-body simulations but not the theory space. However, the system beyond the constrained parameter space will, at some point, enter the relativistic fermion regime or will have too few particles per Hubble volume.

\section{Discussion: Halo Fates} \label{sec:discussion}
Our nonlinear results are only able to follow the first stage of halo formation and not their subsequent evolution.  Firstly, let us explain the reason of behind the smallness of \eqref{eq:mellY}. It turns out that the length scale of the interaction is proportional to the number density of fermions. Indeed, the effective potential for the scalar field in the quartic potential \eqref{eq:Veff} has a minimum at
\begin{align}
\varphi_{\rm min}=-\left(\frac{4y}{\lambda} n_{\psi}\right)^{1/3}\,.
\end{align}
The solution \eqref{eq:ansatzquartic} exhibits the same behaviour. This means that the larger the number of fermions, the larger the vev and the larger the mass of the scalar field. For instance, we have that
\begin{align}
\bar\ell_Y\propto V_{\varphi\varphi}^{-1/2}\propto(\lambda\varphi^2)^{-1/2} \propto(n_\psi y \sqrt{\lambda})^{-1/3}\,,
\end{align}
Thus, although the Yukawa force is much stronger than gravity and creates bound structures in the radiation dominated universe, the typical volume of the basic halos is inversely proportional to the number density. Then, the total mass only depends on $m_\psi$, $y$ and $\lambda$. We note that this density dependent vev for the scalar field also occurs in the so-called chameleon and symmetron models \cite{Brax:2004qh,Burrage:2017qrf,Hinterbichler:2011ca} in the context of dark energy. In the notation of such models, the Yukawa force would be screened for scales $R>(n_\psi y \sqrt{\lambda})^{-1/3}$, were not for the time-oscillations in the Yukawa length scale $\ell$, which render the Yukawa force effectively with a much longer range.

We now comment on various possibilities for the final state of the system, depending on how the evolution of the fermions proceeds.  If the fermions have a small cross-section to standard model particles, the rapid increase in the cores of halos may lead to substantial or complete annihilation of the fermions.  The consequences of such an effect for baryogenesis have been discussed in \cite{Flores:2022oef}.  Note that in this case a different particle is required to make up the dark matter.  If no annihilation occurs, as in the models of asymmetric dark matter~\citep{Petraki:2013wwa,Zurek:2013wia} the coupling to the scalar field also allows the fermions to cool via scalar radiation \cite{bib:Flores2021}.  In this case, the halos first cool to form dark stars with radiation pressure opposing the attractive force.  They can then further cool until the fermion degeneracy pressure becomes the stabilizing force.  If this is also overcome, then they may collapse to form primordial black holes which are decoupled from the scalar field.  The masses of any such collapsed objects depends on how quickly they cool compared to the halo growth rate via mergers.  

If, on the other hand, no radiative behaviour occurs (or is simply inefficient), then the fermions will simply make up $f_\psi$ of the present day matter with inhomogeneities that will simply continue to grow.  The maximum mass of the halos at matter-radiation equality (e.g.~Eq.~\eqref{eq:Mk2}) is around the size of small galaxies, although with strong dependence on $f_\psi$ and $\beta$, and suggests the possibility of using large-scale observations to constrain this scenario.  Let us emphasize that forming such large halos is a very surprising result considering the ``typical" mass scale given by Eq.~\eqref{eq:mellY}.  

While such large halos at matter-radiation equality may be difficult to reconcile with cosmological observations, lighter ones can simply be obtained if the fermions are not all the dark matter, i.e. $f_\psi<1$.  Depending on the compactness and number density of the final objects, we may expect the rest of the dark matter to gravitationally collapse through secondary infall  \cite{bib:Bertschinger1985} with steep density profiles $\rho\propto r^{-9/4}$, similar to what is expected for PBH \cite{bib:Adamek2019,bib:Inman2019}, or via continued hierarchical collapse.  Such dynamics will also be influenced by how the attractive force behaves in the matter era.  Regardless, such processes can then accelerate structure formation in ways that, depending on specific mass scales, may explain or be constrained by high redshift observations such as the EDGES 21 cm signal \cite{bib:Bowman2018}, the excess of heavy galaxies observed by the James Webb Space Telescope \cite{bib:BoylanKolchin2022,bib:Lovell2023}, or the currently unknown formation and growth mechanisms of supermassive black holes \cite{bib:Haiman2013}.  Note that if $f_\psi<1$, then it is necessary to specify the the rest of the dark matter, and some models such as WIMPs may be incompatible with fermions similar to the way they are incompatible with PBHs \cite{bib:Lacki2010,bib:Adamek2019,bib:Carr2021}.

Before concluding, let us briefly compare our results with existing works \cite{Amendola:2017xhl,bib:Savastano2019} and argue that the two system are considerably different. In \cite{Amendola:2017xhl,bib:Savastano2019}, the mass of the fermions is dilatonically coupled to the scalar field, namely in our notation $m_{\rm eff}=m_\psi e^{-\beta\varphi/M_{\rm pl}}$.  The same exponential coupling acts as potential for $\varphi$. For small values of $\varphi$, we have a Yukawa-like interaction where $y=m_\psi\beta/M_{\rm pl}$. However, the exponential running changes the dynamics of the system when compared to the ones studied in this work. With the exponential coupling, the effective mass always decays as $m_{\rm eff}\sim 1/a$ and the energy density of both the fermions and the scalar decays as radiation \cite{Amendola:2017xhl}. Nevertheless, the fermions remain in the non-relativistic regime if they were initially non-relativistic since $m_{\rm eff}/T={\rm constant}$ ($T\propto1/a$ for thermal radiation in an expanding universe). However, the system by itself is never matter dominated unless there is backreaction from the non-linear structures or one considers, e.g., a bare mass for $m_\psi$. We also find that, in the dilatonic coupling case, the effective scalar field mass $M_\varphi^2$ decays with the scale factor and all perturbations on subhorizon scales ($k\gg{\cal H}$) grow. This growth is a power-law with $\delta_\psi\sim a^p$ and $p\sim1.62$ \cite{Amendola:2017xhl}, which is not as fast as the exponential growth of this work.

\section{Conclusions} \label{sec:conclusions}

We have computed the dynamics of a realistic interacting dark sector composed of just one scalar and one family of fermion particles via a Yukawa coupling.  We showed that including a potential for the scalar field allows for the fermions to remain non-relativistic behaving as dark matter with an additional attractive force. We then focused on a quartic potential for the scalar field, which is motivated by renormalizable theories and also convenient for the simulations. Confirming previous studies \cite{bib:Flores2021}, we find that this attractive force leads to substantial growth in the fermions allowing for the possibility of nonlinear halo formation.  We also find that the specific type of coupling substantially changes the growth of perturbations and can lead to the formation of halos much larger than has previously been assumed in simplified setups. These halos could become dark stars, primordial black holes, galaxy-sized halos at matter-radiation equality, or end up annihilating. Such halo formation might also be accompanied by gravitational waves \cite{Flores:2022uzt}, cold electroweak baryogenesis \cite{Flores:2022oef} or magnetogenesis \cite{Durrer:2022cja}. 

While our results have demonstrated the nonlinear halo formation in the scalar-fermion dark sector, they are also limited in some ways.  On the theory side, we have assumed the radiation era throughout as we expected halos to be small in the radiation era.  Determining how the system behaves into the matter epoch is of substantial interest from the perspective of large-scale structure.  

We have also been limited by various aspects of the numerical simulations beyond the lack of radiative physics.  Firstly, the simulations are fully Newtonian and in comoving coordinates, requiring the fermions to be non-relativistic at all times, that the growth of perturbations occurs on subhorizon scales and that $\ell$ is on average constant.  These requirements make other potentials such as the quadratic one difficult to simulate.  We have also been substantially limited by resolution.  On small-scales, our use of just a particle-mesh calculation does not allow for detailed investigation of the halo interiors and less concentrated halos may interact with the finite value of $\ell$ in unexpected ways.  Including a subgrid pairwise force would circumvent this issue and is important to include in the future.  On large-scales, we lack the ability to follow the $\sqrt{k\bar\ell}$ scaling limiting the maximum redshift and mass of the halos in our simulation.  This regime could however be studied using the asymptotic high-frequency scaling alone.

Let us conclude by highlighting that the scalar-fermion system we have considered is {\it simple}. The  long-range forces between matter particles can arise in models with supersymmetry~\cite{Flores:2021jas}, as well as other dark matter models~\cite{Gradwohl:1992ue,Kesden:2009bb,Carroll:2009dw,Archidiacono:2022iuu}.  We therefore expect the phenomenology we have identified to be rather common in expanded dark sectors in the radiation epoch.  Moving from such a phenomenological discussion of final states - radiation, stars, black holes, or halos - to predictive modelling, as well as generalizing to various other interacting dark sectors, are therefore important steps towards understanding the dynamics of the early Universe. 

\section*{Acknowledgements}
We thank Elisa G. M. Ferreira, Marcos Flores, Lauren Pearce, Javier Rubio, Volodymyr Takhistov, Edoardo Vitagliano, Christof Wetterich and Graham White for valuable discussions.  GD is supported by the DFG under the Emmy-Noether program grant no. DO 2574/1-1, project number 496592360.  This work  was supported  by  the World Premier International Research Center Initiative (WPI),  MEXT,  Japan and by Japan Society for the Promotion of Science (JSPS) KAKENHI grant No. JP20H05853.  A.K. was also  supported  by the U.S. Department of Energy (DOE) Grant No.  DE-SC0009937.  
This work made use of {\sc NumPy} \citep{bib:Harris2020}, {\sc SciPy} \citep{bib:Virtanen2020}, {\sc Matplotlib} \citep{bib:Hunter2007} and NASA's Astrophysics Data System Bibliographic Services.

\appendix

\section{Parameters and definitions \label{app:elliptic}}
In this appendix we summarize the meaning of main parameters and definitions used throughout the text.

\subsection{Jacobi elliptic functions\label{app:jacobi}}
Here we define the Jacobian elliptic functions. The Jacobian elliptic functions are the inverse of the elliptic integrals, explicitly given by 
\begin{align}
\zeta=F_\alpha(\phi)=\int_0^\phi\frac{d\theta}{\sqrt{1-\alpha^2\sin^2\theta}}\,.
\end{align}
One then defines
\begin{align}
{\rm Cn}_\alpha(\zeta)\equiv\cos\phi\,,
\end{align}
from which is clear that ${\rm Cn}_\alpha(\zeta)$ is periodic. The period $\Delta x$ defined by ${\rm Cn}_\alpha(\zeta)={\rm Cn}_\alpha(\zeta+\Delta \zeta)$ is given by
\begin{align}
\Delta \zeta=F_\alpha(2\pi)=4K_\alpha\,,
\end{align}
where $K_\alpha$ is the complete elliptic integral of the first kind.
For the case of \S\ref{subsec:quartic}, that is $\alpha=2^{-3/2}(\sqrt{3}-1)$, we find the period can also be expressed in terms of Gamma functions as
\begin{align}
K_\alpha= 3^{1/4}\sqrt{\pi}\frac{\Gamma[7/6]}{\Gamma[2/3]}\approx 1.6\,.
\end{align}

\subsection{Parameters and Variables\label{app:definitions}}

In table \ref{tab:1} we provide a list of the main parameters and variables.  As an introduction section, \S\ref{sec:motivation} does not make the same approximations (e.g.~$\beta^2\gg1$ and $a\ll a_{\rm eq}$) as the remainder of the text and so the notation differs.  Nonetheless, if those approximations are made then the transformation $6\alpha x\rightarrow \bar{A}_{k\bar\ell} s$ (see below \eqref{eq:approx12}) yield the same equations.  Lastly, many parameters are with respect to matter-radiation equality, defined as when $\rho_r=\rho_m+\rho_\psi$ (i.e.~excluding $\varphi$), and denoted with subscript ``${\rm eq}$''.  In particular, $k_{\rm eq}=a_{\rm eq}H_{\rm eq}$ is the comoving horizon size at that time.

\begin{table}[htp]
\begin{tabularx}{\columnwidth}{ll}
\multicolumn{2}{>{}l}{\fontsize{11pt}{11pt} \textsc{Scalar Field, $\varphi$}}\\
\toprule
Bare mass & $m_\varphi$\\
Quartic coupling & $\lambda$\\
Potential & $V=m_\varphi^2\varphi^2/2+\lambda\varphi^4/4+...$\\
Effective mass & $M_\varphi=a \sqrt{V_{\varphi\varphi}}$\\
Dimensionless effective mass & $\mu\propto M_\varphi / k_{\rm eq}$\\
Perturbation & $\delta\varphi$\\ \\
\multicolumn{2}{>{}l}{\fontsize{11pt}{11pt} \textsc{Fermions, $\psi$}}\\
\toprule
Bare mass & $m_\psi$\\
Number density & $n_\psi$\\
Energy density & $\rho_\psi=m_{\rm eff} n_\psi$\\
Matter fraction in $\psi$ & $f_\psi=\rho_\psi/(\rho_m+\rho_\psi)$\\ 
Perturbation & $\delta_\psi = \delta n_\psi/n_\psi$ \\ \\
\multicolumn{2}{>{}l}{\fontsize{11pt}{11pt} \textsc{Yukawa Interactions}}\\
\toprule
Yukawa coupling & $y$\\
Interaction strength & $\beta=y M_{\rm pl}/m_\psi$\\
Effective $\psi$ mass & $m_{\rm eff}=m_\psi+y\varphi$\\
Effective $\varphi$ potential & $V_{\rm eff}=V+{\rm sign}[m_{\rm eff}]y\varphi n_\psi$\\
Long-range potential & $\phi_Y=y\delta\varphi/m_{\rm eff}$ \\
Yukawa length scale & $\ell=M_\varphi^{-1}\propto (k_{\rm eq}\mu)^{-1}$ \\
Effective scale factor & $s=12f_\psi\beta^2 a/a_{\rm eq}$ \\
Effective frequency (quartic) & $\omega$ ($\omega s\propto \mu \,a/a_{\rm eq}$) \\ \\
\multicolumn{2}{>{}l}{\fontsize{11pt}{11pt} \textsc{Gravitational Interactions}}\\
\toprule
Scale factor & $a$ \\
Hubble parameter & $H=a^{-1}da/dt = \mathcal{H}/a$ \\
Metric potentials & $\phi_G=\Phi\simeq-\Psi$\\
Matter density, w/out $\psi$, $\varphi$ & $\rho_m$\\
Radiation density, w/out $\psi$, $\varphi$ & $\rho_r$ \\
Perturbations & $\delta_r=\delta \rho_r/\rho_r$, $\delta_m =\delta\rho_m/\rho_m$
\end{tabularx}
\caption{List of basic parameters of our model. In the main text quantities with an upper ``bar'' are constant and refer to the amplitude of the unbarred variable. For example, $\bar\ell$ is the amplitude of $\ell$. \label{tab:1}}
\end{table}

\section{Detailed conditions for non-relativistic fermions \label{app:detailedconditions}}

In this appendix we provide the limits used to draw figures \ref{fig:conditiongeneral}, \ref{fig:conditionmphi2} and \ref{fig:conditionlambda2}.

\subsection{Massless case}

For the massless case we have that conditions \ref{iia},\ref{iib} and \ref{iii} respectively yield
\begin{align}\label{eq:boundonbetaml1}
    \beta\ll 4\times 10^9f_\psi^{-1/2}\left(\frac{g_{*}(T)}{106.75}\right)^{1/8}\left(\frac{m_\psi}{10^{10}\,{\rm GeV}}\right)^{1/2}\,,
\end{align}
\begin{align}\label{eq:boundonbetaml2}
    \beta\ll 2\times 10^{12}f_\psi^{-2/3}\left(\frac{m_\psi}{10^{10}\,{\rm GeV}}\right)^{2/3}\,,
\end{align}
and
\begin{align}\label{eq:boundonbetaml3}
    \beta\ll 3\times 10^{10}f_\psi^{-1/3}\left(\frac{m_\psi}{10^{10}\,{\rm GeV}}\right)^{-1/6}\,.
\end{align}

\subsection{Quadratic potential}

For the case of a quadratic potential, we find the following. On one hand, for condition \ref{iia} we require that $T_1\ll m_\psi$, where $T_1$ temperature of the radiation fluid at that the time $\tau_1$ and which reads
    \begin{align}
    \frac{T_1}{M_{\rm pl}}\approx 0.33\sqrt{\frac{m_\varphi}{M_{\rm pl}}}\left(\frac{g_{*}(T)}{106.75}\right)^{-1/4}\,.
    \end{align}
    Then, the condition $T_1\ll m_\psi$ yields
    \begin{align}\label{eq:boundquadratic2}
    m_\varphi\ll 370\,{\rm GeV}\left(\frac{g_{*}(T)}{106.75}\right)^{1/2}\,.
    \end{align}
    To have non-zero parameter space between the bounds \eqref{eq:boundquadratic} and \eqref{eq:boundquadratic2}, we obtain the following upper bound on $\beta$:
    \begin{align}\label{eq:boundonbeta}
    \beta\ll 5\times 10^9f_\psi^{-1/2}\left(\frac{g_{*}(T)}{106.75}\right)^{1/8}\left(\frac{m_\psi}{10^{10}\,{\rm GeV}}\right)^{1/2}\,.
    \end{align}

On the other hand, for condition \ref{iib} we require that $m_\psi^3\gg 3\pi^2 n_\psi$ evaluated at $\xi_1$ which yields
    \begin{align}
    m_\varphi\ll 3.4\times 10^{13}\,{\rm GeV}f_\psi^{-2/3}\left(\frac{m_\psi}{10^{10}\,{\rm GeV}}\right)^{8/3}\,.
    \end{align}
    We see that the case of degenerated non-relativistic fermions is less restrictive in the value of $m_\varphi$ and the upper bound on $\beta$ now reads
    \begin{align}\label{eq:boundonbetadegenerate}
    \beta\ll 3\times 10^{12}f_\psi^{-2/3}\left(\frac{m_\psi}{10^{10}\,{\rm GeV}}\right)^{2/3}\,.
    \end{align}
    Lastly, condition \ref{iii} also evaluated at $\xi_1$ implies
    \begin{align}
    m_\varphi\ll 2\times 10^{6}\,{\rm GeV}f_\psi^{2/3}\left(\frac{m_\psi}{10^{10}\,{\rm GeV}}\right)^{-2/3}\,.
    \end{align}

\subsection{Quartic potential}

We proceed as in the previous subsection but for the quartic potential. First, for condition \ref{iia}, the temperature of the radiation fluid at $a_{\rm max}$ is given by
\begin{align}
\frac{T(a_{\rm max})}{M_{\rm pl}}\approx 7\times10^{-29}\mu\left(\frac{g_{*}(T)}{106.75}\right)^{-1/4}\,.
\end{align}
Then, by imposing $T(a_{\rm max})\ll m_\psi$ we obtain
\begin{align}\label{eq:boundquartic2}
\lambda<\frac{1.4\times 10^{7}}{f_\psi^2\beta^2}\left(\frac{g_{*}(T)}{106.75}\right)^{3/2}\left(\frac{m_\psi}{10^{10}\,{\rm GeV}}\right)^{6}\,.
\end{align}
By saturating the inequalities \eqref{eq:boundquartic1} and \eqref{eq:boundquartic2} we also obtain an upper bound on $\beta$ given by \eqref{eq:boundonbeta}. 

For condition \ref{iib}, we require $m_\psi\ll 3\pi^2n_\psi(a_{\rm max})$, which yields
\begin{align}\label{eq:boundquartic3}
\lambda<\frac{1.1\times 10^{40}}{f_\psi^4\beta^2}\left(\frac{m_\psi}{10^{10}\,{\rm GeV}}\right)^{8}\,,
\end{align}
and again the resulting upper bound on $\beta$ by saturating the parameter space coincides with the one in the quadratic case \eqref{eq:boundonbetadegenerate}.
Lastly, the condition \ref{iii}, that requires more than one particle per Hubble volume, imposes
\begin{align}\label{eq:boundquartic4}
\lambda<\frac{3\times 10^{18}}{\beta^2}\left(\frac{m_\psi}{10^{10}\,{\rm GeV}}\right)^{-2}\,.
\end{align}

\section{Einstein equations\label{app:einsteinequations}}

In this appendix we explicitly write the general equations used in this paper. We start with Einstein equations which are given by
\begin{align}\label{eq:EE}
M_{\rm pl}^2 G_{\mu\nu}&=T_{\psi,\mu\nu}+T_{r,\mu\nu}+T_{m,\mu\nu}\nonumber\\&
+\partial_\mu\varphi\partial_\nu\varphi-\frac{1}{2}g_{\mu\nu}\left(\partial_\alpha\varphi\partial^\alpha\varphi+2V(\varphi)\right)\,,
\end{align}
where the subscript $r$ refers to radiation and $T_{Q,\mu\nu}$ with $Q=\{r,m,\psi\}$ is the energy-momentum tensor of a perfectly fluid, explicitly given by
\begin{align}\label{eq:Trmunu}
T_{Q,\mu\nu}=\left(\rho_Q+P_Q\right)u_{Q,\mu}u_{Q,\nu}+P_Qg_{\mu\nu}\,,
\end{align}
and $u_{Q,\mu}$ are the fluid's 4-velocity. Second, the energy conservation of the Fermi gas is given by
\begin{align}\label{eq:energycons}
u_\psi^\mu\nabla_\mu\rho_\psi+\left(\rho_\psi+P_\psi\right)\nabla_\mu u_\psi^\mu+\left(\frac{\partial P_\psi}{\partial\varphi}\right)_{T,\mu}u_\psi^\mu\nabla_\mu\varphi=0\,.
\end{align}
Then, we use the Bianchi identities and the conservation of the energy momentum tensor for radiation, to derive the Klein-Gordon equation which reads
\begin{align}\label{eq:KGapp}
\nabla_\nu\nabla^\nu\varphi-V_\varphi-\frac{m_{\rm eff}}{|m_{\rm eff}|}{y}n_\psi=0\,.
\end{align}
In a similar manner, we find that the equation for the velocity is given by
\begin{align}\label{eq:velocity}
& u_\psi^\nu\nabla_\nu u_{\psi,\mu}+\left(\delta_{\mu}^\nu-u_{\psi,\mu}u^\nu_{\psi}\right)\frac{1}{\rho_\psi}\frac{y}{m_{\rm eff}}\nabla_\nu\varphi=0\,.
\end{align}
Lastly, we have the fermion number density conservation, namely
\begin{align}\label{eq:conservation}
\nabla_\mu(n_\psi u_\psi^\mu)=0\,.
\end{align}

\subsection{Cosmological perturbations on subhorizon scales\label{app:cosmopertu}}

We also present here the main equations for cosmological perturbations that are used in the text. We perturb the FLRW metric in the shear-free gauge, in which the line element reads
\begin{align}
ds^2=a^2(-(1+2\Psi)d\eta^2+(1+2\Phi)dx^2)\,.
\end{align}
The energy densities are expanded as $\rho\to \rho+\delta\rho$ and the velocities as $u^\mu=a^{-1}(1-\Phi,v^i)$ for both the radiation and fermion fluids. For the scalar field we take $\varphi\to \varphi+\delta\varphi$. With this prescription, the $00$ component of Einstein equations, which is related to the Poisson equation, is given by
\begin{align}
2\frac{ k^2 }{{\cal H}^2}\Phi&=\frac{a^2 \rho_r}{{\cal H}^2
   M_{\rm pl}^2}\delta_r+\frac{a^2 \rho_r}{{\cal H}^2
   M_{\rm pl}^2}\delta_m +\frac{a^2  \rho_\psi }{{\cal H}^2 M_{\rm pl}^2}\delta_\psi\nonumber\\&+\frac{a^2 \rho_\psi }{{\cal H}^2 M_{\rm pl}^2}\frac{y}{m_{\rm eff}}\delta\varphi+\frac{1}{M_{\rm pl}^2}\frac{d\varphi}{dN}\frac{d\delta \varphi}{dN}+\frac{a^2 V_\varphi}{{\cal H}^2
   M_{\rm pl}^2}\delta\varphi\,.
\end{align}
Then we have the Klein gordon equation:
\begin{align}
\frac{d^2\delta\varphi}{dN^2}&+ \frac{d\delta\varphi}{dN}+ 
   \frac{k^2}{{\cal H}^2}\delta\varphi+\frac{y }{ m_{\rm eff}}\frac{a^2  \rho_\psi}{{\cal H}^2} \delta_\psi=0\,.
\end{align}
The energy density and momentum conservation for fermions:

\begin{align}
\frac{d\delta_\psi}{dN}-\frac{k^2}{{\cal H}} v_\psi
   =0\,,
\end{align}

\begin{align}
\frac{dv_\psi}{dN} +v_\psi-\frac{1}{\cal H}
    \Phi+\frac{1}{\cal H}\frac{y}{m_{\rm eff}} \delta\varphi  =0\,.
\end{align}
And the energy density and momentum conservation for radiation:
\begin{align}\label{eq:drapp}
\frac{d\delta_r}{dN}-\frac{4}{3}\frac{k^2}{{\cal H}} v_r=0\,,
\end{align}
\begin{align}
 \frac{dv_r}{dN}+\frac{1}{{\cal H}}\left(\frac{1}{4}\delta_r -\Phi\right)=0\,.
\end{align}
If there is an additional dark matter component which behaves like dust, we also have:
\begin{align}\label{eq:dmapp}
\frac{d\delta_m}{dN}-\frac{k^2}{{\cal H}} v_m
   =0\,,
\end{align}
\begin{align}
\frac{dv_m}{dN} +v_m-\frac{1}{\cal H}
    \Phi  =0\,.
\end{align}

\subsection{Newtonian Approximation}
\label{subsubsec:NApp}
For completeness we derive the equations in the Newtonian regime, which consists in keeping the linear order of metric perturbations but allowing for the non-linear evolution of the non-relativistic matter fields. This is justified since on subhorizon scales the gravitational potential is suppressed with respect to the density fluctuations. Here we do the same perturbative expansion as in appendix \ref{app:cosmopertu} but we also include the velocity perturbation of the fermion fluid. With this prescription, the relevant equations in real space and in terms of e-folds are given by
\begin{align}
    &\frac{d{\delta_\psi}}{dN}+\frac{1}{aH}\vec{\nabla}\left[(1+\delta_\psi)\vec{v}_\psi\right]=0\,,\\
    &\frac{d\vec{v}_\psi}{dN}+\left(1+\frac{d\ln m_{\rm eff}}{dN}\right)\vec{v}_\psi\nonumber\\&\phantom{\frac{d\vec{v}_\psi}{dN}}+\frac{1}{aH}(\vec{v}_\psi\cdot\vec{\nabla})\vec{v}_\psi+\frac{1}{aH}\vec{\nabla}\phi_Y=0 \,,\\
    &(\nabla^2-\ell^{-2})\phi_Y=a^2{\beta^2}M_{\rm pl}^{-2}\rho_\psi\delta_\psi\label{eq:phiY} \,,
\end{align}
where $\phi_Y$ is  the effective potential due to the Yukawa interaction related to the scalar field by \eqref{eq:phidefinition} and $\ell^{-2}=M_\varphi^2=a^2V_{\varphi\varphi}$. These equations can be recast in a more convenient form with the following redefinitions:
\begin{align}
v_\psi=\left(\frac{\sqrt{2} a}{a_{\rm eq}}\right)^{-1}V_\psi\quad&,\quad\phi_Y=\left(\frac{\sqrt{2} a}{a_{\rm eq}}\right)^{-2}\phi_N\,,\nonumber\\
 \vec{\nabla}=a_{\rm eq}H_{\rm eq}\vec{\tilde\nabla}\quad&{,}\quad \ell=a^{-1}_{\rm eq}H^{-1}_{\rm eq}\,\tilde \ell\,.
\end{align}
We also introduce a new time variable by
\begin{align}
s=12f_\psi\beta^2 \frac{a}{a_{\rm eq}}\,.
\end{align}
Then, we have that $dN=d\ln s$. In this notation we obtain
\begin{align}\label{eq:newtonian}
    &\frac{d{\delta_\psi}}{d\ln s}+\vec{\tilde\nabla}\left[(1+\delta_\psi)\vec{V}_\psi\right]=0\,, \\
    &\frac{d{\vec{V}}_\psi}{d\ln s}+\frac{d\ln m_{\rm eff}}{d\ln s}\vec{V}_\psi+(\vec{V}_\psi\cdot\vec{\tilde \nabla})\vec{V}_\psi+\vec{\tilde\nabla}\phi_N=0\,, \\
    &(\tilde \nabla^2-\tilde \ell^{-2})\phi_N=\frac{s}{4}f_\psi\delta_\psi\,,
\end{align}
where we used \eqref{eq:npsitoHeq} to write $n_{\psi,\rm eq}$ in terms of $H_{\rm eq}$. It is important to note that the equations above are independent of $H_{\rm eq}$ and $\beta$. Thus, we may solve them in general and later translate the result to the physical variables.

\section{Choice of \texorpdfstring{$\bar{\ell}$}{barell}}
    \label{app:lbar}
    In this appendix, we consider how the value of $\bar{\ell}$ ($n_\ell$) relative to the simulation size $L$ ($n_c$) affects our calculation.  First of all, we consider the calculation of the force in Eq.~\eqref{eq:fN}.  We run a set of pairwise force calculations between two particles as a function of their separation $r_s$ and the value of $n_\ell$.  We show the results in Fig.~\ref{fig:force_ny}.  Despite the modification to include $\bar\ell$, the particle-mesh calculation performs well for particles separated by more than a few grid cells.  As separations approach $n_c/2$ the periodic boundary conditions truncate the force, rather than $n_\ell$.

    While the physical value of $\ell$ cannot affect the results of the simulation, the numerical value $n_\ell$ certainly can via artifacts associated with finite resolution.  If the value is small and approaches the grid size $n_\ell\sim1$, then particles will not feel any force.  We would like then, to choose $n_\ell$ fairly large.  However, if it is too large then our results will be affected by the simulations periodic boundary conditions which will truncate the force law rather than the Yukawa scale.  This is particularly troublesome for our scale free initial conditions, as nonlinear coupling between modes within the simulation volume and those larger than the box would not be resolved.  This motivates a choice of $n_\ell$ fairly small.
    
    Fortunately, the invariance to a physical value of $\ell$ allows for a useful test, analogous to scale-free tests for large-scale structure simulations \citep{bib:Joyce2021}, as converged results should be insensitive to a specific choice of $n_\ell$ provided they are rescaled consistent with their dimensions.  We therefore run a set of simulations with constant $m_s$ and $\ell_s$, but varying the choice of $n_\ell$.  We note that even though the initial density field is the same, the force law effectively differs between the simulations and so the final results are only statistically equivalent.  

\begin{figure}
        \centering
        \includegraphics[width=0.9\columnwidth]{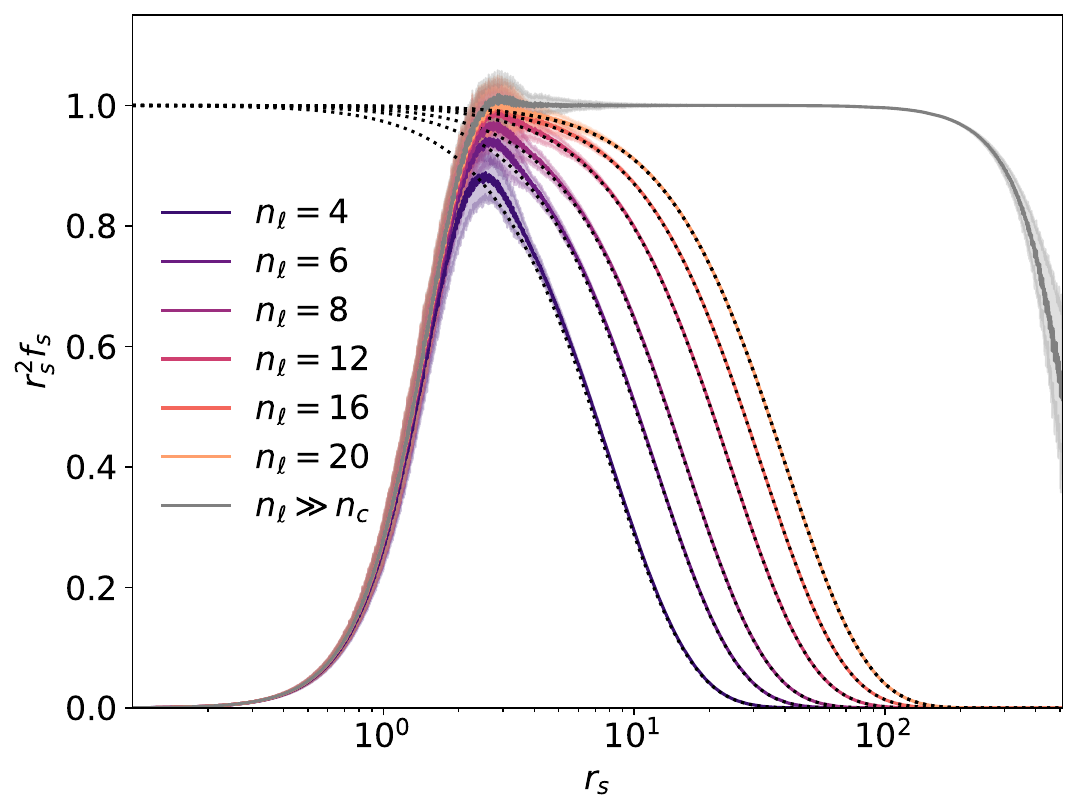}
        \caption{Pairwise computation of the Yukawa force $f_s$ for various choices of $n_\ell$.  $r_s$ is the particle separation (specified in grid cells, with $n_c=1024$), and we normalized $r_s^2f_s$ to $1$ for $r_s\ll n_\ell$.  $1\sigma$ fluctuations are computed by repeating the calculation $100$ times.}
        \label{fig:force_ny}
    \end{figure}
    
    \begin{figure}
        \centering
        \includegraphics[width=0.9\columnwidth]{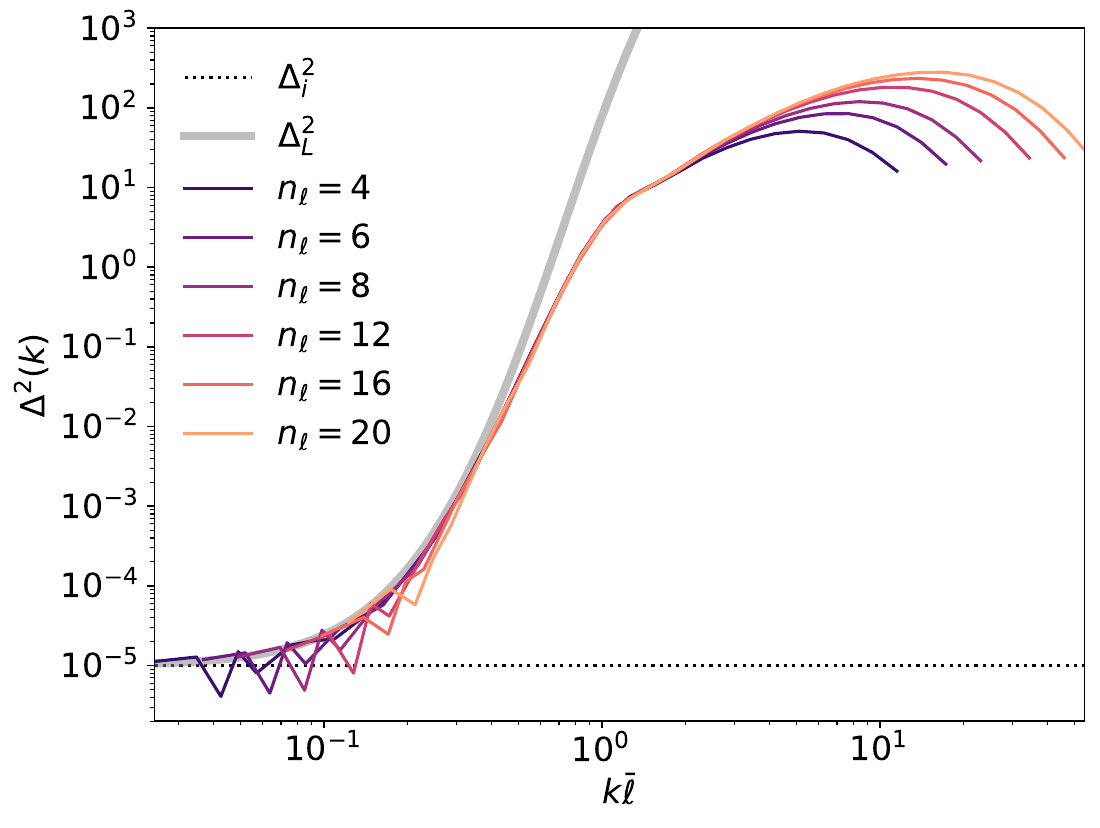}
        \caption{Dimensionless power spectra at $s=200$ for constant $m_{\rm eff}$ and $\ell$ but different choices of $n_\ell$ relative to the grid size, $n_c=1024$.}
        \label{fig:power_ny}
    \end{figure}
    
    We show our results at $s=200$ in Fig.~\ref{fig:power_ny}.  We see that unlike the linear solution which grows arbitrarily, halos virialize with more typical densities.  We see that, as expected, larger values of $n_\ell$ better probe small scales, while smaller values are able to resolve the constant large scales.  In addition, we also consider the halo mass function of these simulations, shown in Fig.~\ref{fig:hmf_ny}.  Here we do not see as good agreement upon rescaling, and simulations with $n_\ell\leq8$ having fairly different mass functions.  We also see a decrease in heavy halos when $n_\ell=20$ (and also a little for $16$), which is indicative of having too small a volume for such large halos.  Fortunately, $n_\ell=12$ and $n_\ell=16$ appear reasonably converged in both the power spectrum and the halo mass function.  

    \begin{figure}
        \centering
        \includegraphics[width=0.9\columnwidth]{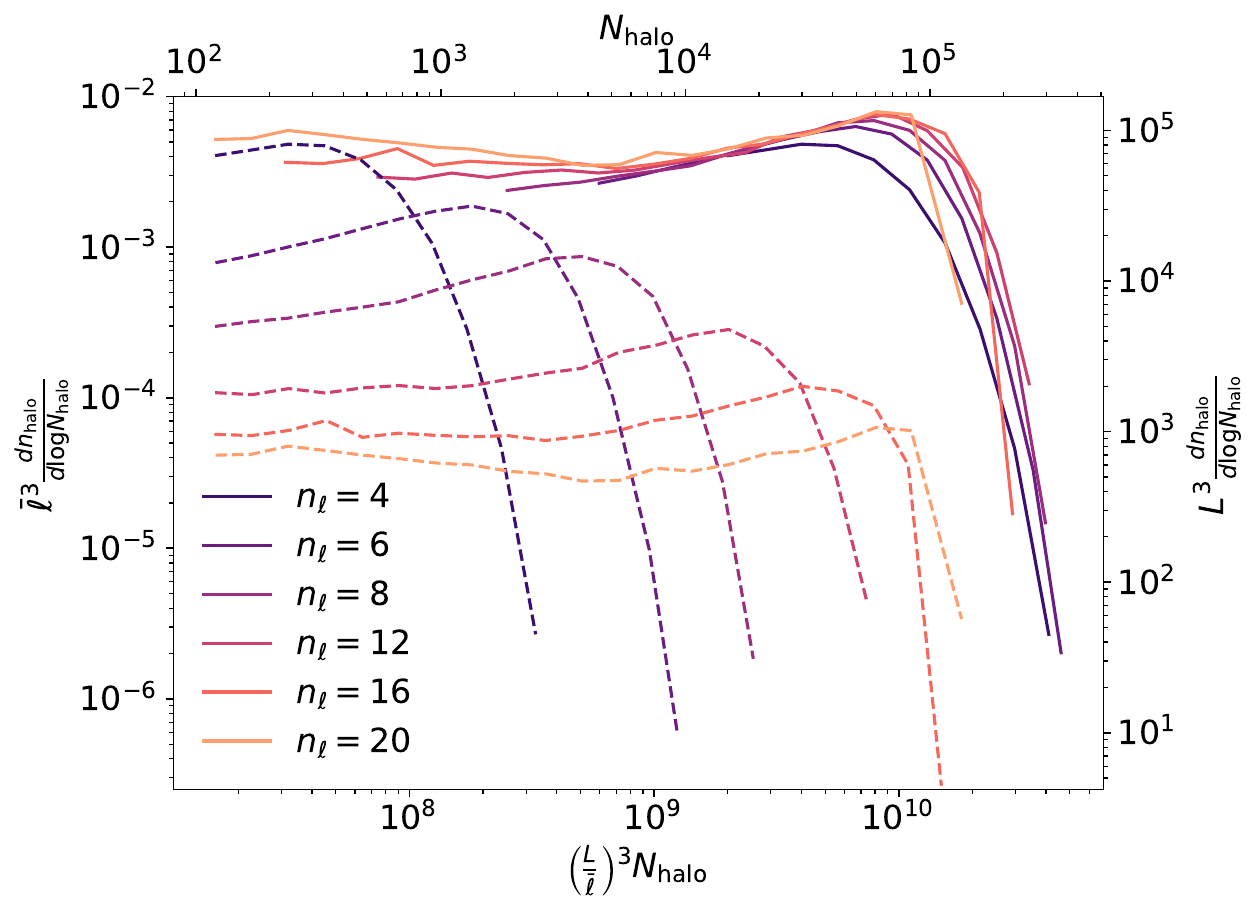}
        \caption{Halo mass function at $s=200$ as a function of the number of N-body particles in the halo $N_{\rm halo}$.  The solid curves show the rescaled mass functions which should be equivalent outside of numerical artifacts.  The dashed curves are unscaled and correspond to the upper x-axis and right y-axis.}
        \label{fig:hmf_ny}
    \end{figure} 

\begin{figure}
        \centering
        \includegraphics[width=0.9\columnwidth]{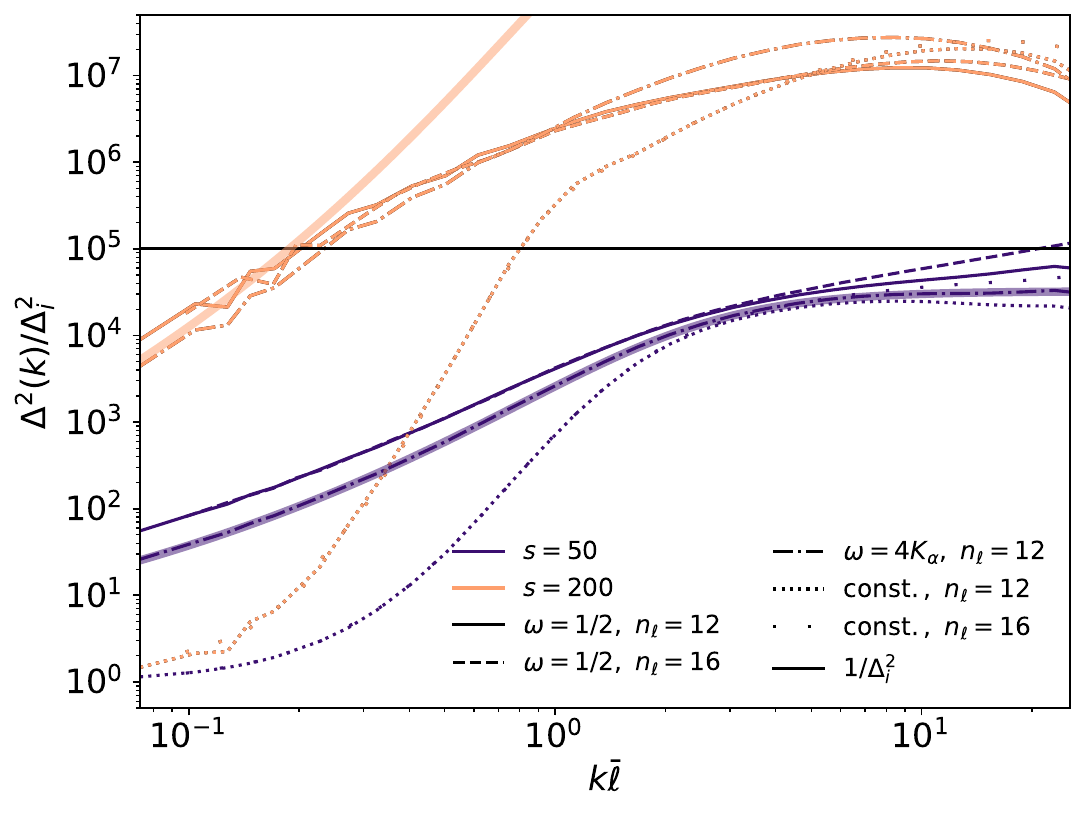}
        \caption{Dimensionless power spectra at $s=50$ and $200$ for various values of $\omega$ and $n_\ell$.  The bands show the linear result using the high-frequency approximation.}
        \label{fig:power_LF_ny}
    \end{figure}
     
     When including time dependence of $\ell$ there is an additional affect to consider which is that the minimum value of $\ell$ is $\sim\bar\ell/2$.  For the high-frequency case, we expect the simulation to match the effective force law in \eqref{eq:fit}; however, for the low frequency case substantial amounts of the simulation time is spent around this minimum which may amplify force artifacts. We do a test by running a low frequency simulation with $n_\ell=16$.  We show the power spectra at before the onset of nonlinearity ($s=50$) and after ($s=200$) in Fig.~\ref{fig:power_LF_ny}.  
     We find that the two choices of $n_\ell$ lead to very similar results; however, we also find that the low frequency case has lower power in the nonlinear regime of the simulation than the high frequency case.  It is not clear why this may occur, but could be indicative of a lack of force at $\ell\sim\bar\ell/2$.  In general, without a subgrid force we expect that halos are less concentrated and the power spectrum to be lower regardless of our choices of $n_\ell$ or $\omega$.  This may have unexpected consequences, as the limited range of the force means that the distribution of matter in the halo, and not just the halo mass, matters \cite{bib:Martino2009}.  While we may expect the long-range periods to ameliorate this effect, it remains to be seen whether there are  differences in the halo distribution when subgrid forces are included.

\bibliography{thebib}
    
\end{document}